\documentclass[apjl]{emulateapj}
\usepackage{natbib}
\usepackage{amsmath}
\usepackage{graphicx}
\usepackage{epsfig}
 
\usepackage{amsfonts}

\def\Dwa{$\,$\uppercase\expandafter{\romannumeral5}$\,$}

\def\sless{\lower2pt\hbox{$\buildrel {\scriptstyle <}
   \over {\scriptstyle\sim}$}}

\def\sgreat{\lower2pt\hbox{$\buildrel {\scriptstyle >}
   \over {\scriptstyle\sim}$}}

\begin{document}


\title{Should One Use the Ray-by-Ray Approximation in Core-Collapse Supernova Simulations?} 
\author{M. Aaron Skinner\altaffilmark{1}, Adam Burrows\altaffilmark{1}, \& Joshua C. Dolence\altaffilmark{2}}
\altaffiltext{1}{Department of Astrophysical Sciences, Princeton University, Princeton, NJ 08544; 
burrows,askinner@astro.princeton.edu}
\altaffiltext{2}{CCS-2, Los Alamos National Laboratory, P.O. Box 1663 
Los Alamos, NM 87545; jdolence@lanl.gov}

\begin{abstract}
We perform the first self-consistent, time-dependent, multi-group
calculations in two dimensions (2D) to address the consequences of using
the ray-by-ray+ transport simplification in core-collapse supernova
simulations. Such a dimensional reduction is employed by many
researchers to facilitate their resource-intensive calculations. Our new
code (F{\sc{ornax}}) implements multi-D transport, and can, by
zeroing out transverse flux terms, emulate the ray-by-ray+ scheme. 
Using the same microphysics, initial models, resolution, 
and code, we compare the results of simulating 12-, 15-, 20-, and 
25-M$_{\odot}$ progenitor models using these two transport
methods. Our findings call into question the wisdom of the pervasive use
of the ray-by-ray+ approach.  Employing it leads to maximum
post-bounce/pre-explosion shock radii that are almost universally larger
by tens of kilometers than those derived using the more
accurate scheme, typically leaving the post-bounce matter less bound and artificially
more ``explodable." {In fact, for our 25-M$_{\odot}$ progenitor,
the ray-by-ray+ model explodes, while the corresponding multi-D transport model does not.
Therefore, in two dimensions the combination of ray-by-ray+ with the 
axial sloshing hydrodynamics that is a feature of 2D supernova dynamics can result 
in quantitatively, and perhaps qualitatively, incorrect results.}
\end{abstract}


\section{Introduction}
\label{intro}

Despite years of tangible advance, the core-collapse supernova problem has persisted 
for half a century without a definitive solution.  While there has been much conceptual progress
during this time concerning the relevant nuclear, particle, and neutrino physics, and while the astrophysical context
and goals of theory have been greatly sharpened and refined, the mechanism of explosion that leads to the measured
explosion energies, nucleosynthesis, debris morphology, and residues (neutron stars or stellar-mass black holes)
has not been unambiguously determined. Neutrinos are copiously produced and radiated at the high 
densities and temperatures achieved in collapse, and it is widely believed that neutrino heating plays a pivotal role
in igniting the explosion (Colgate \& White 1966; Bethe \& Wilson 1985).  However, the hydrodynamic instabilities 
and turbulence that attend collapse, bounce, and proto-neutron star mantle dynamics alter the critical condition for 
explosion (Burrows \& Goshy 1993; Murphy \& Burrows 2008; Pejcha \& Thompson 2012) 
by introducing turbulent stress and enhancing the efficiency of the neutrino-matter 
coupling and must be simulated in detail to properly capture such (and related) effects. 
The multi-dimensional complexity of the hydrodynamics and neutrino radiative transfer seem to be central 
to the mechanism itself $-$ by a wide margin, spherical models don't generally explode. This requires
the solution of the coupled partial-differential equations of multi-dimensional hydrodynamics and radiative transfer,
which classically is a seven-dimensional problem (time, three space, and three momentum space for each neutrino species).  
If performed with acceptable resolution, multi-angle transport at every three-dimensional position 
at every timestep can not be done using current supercomputers.  Hence, various simplifications, mostly 
in the radiation sector, have been introduced by the different groups attacking the core-collapse mechanism problem.
In this paper, we address the accuracy of one such approximation, the so-called ``ray-by-ray" transport method.
While our group does not use it (Burrows et al. 2006,2007; Dolence, Burrows, \& Zhang 2015), 
it is being employed by most groups currently publishing explosions, and the fidelity with which it handles
multi-dimensional transport, particularly in two spatial dimensions, has been called into question 
(Burrows 2013; Dolence, Burrows, \& Zhang 2015; Sumiyoshi et al. 2015). 

Using their CHIMERA code, the ORNL group has obtained tepid explosions in 2D (Bruenn et~al. 2013,2014) 
and in 3D (Lentz et al. 2015). The Garching group, using the CoCoNuT hydrodynamics code (M\"uller et~al.
2012ab; M\"uller \& Janka 2014) in combination with the VERTEX transport
solver, or their earlier VERTEX-PROMETHEUS code (Buras et~al. 2006), has also obtained tepid explosions in 2D,
and in 3D only when altering the physics (Melson et al. 2015). Suwa et al. (2014), Takiwaki et al. (2012), 
and Iwakami, Nagakura, \& Yamada (2014) neglect $\nu_{\mu}$ and $\nu_{\tau}$ neutrinos and use the simplifying IDSA 
technique (Liebend\"orfer et al. 2009); they also obtain explosions in 2D and have explored 
3D simulations at low resolution.  

However, none of these groups performs multi-dimensional transport, but all use multiple one-dimensional 
spherical transport solves in lieu of more accurate multi-dimensional solutions in the so-called 
``ray-by-ray" approach (Buras et al. 2003,2006; Burrows, Hayes, \& Fryxell 1995). 
One-dimensional transport using thermodynamic
profiles along a single radial ray is calculated and coupled only locally
to the spherical hydro solution of that ray.  This is done for all, or a subset,
of the radial rays to approximate the radiation field and its evolution.
Such a dimenisional-reduction technique is employed to facilitate 
multi-processor scaling and reduce computational overhead. However, transverse 
radiation fluxes and lateral transport are ignored, which in reality would significantly 
smooth out the radiation field.  The relative smoothness of a radiation field, even when the 
background thermal field is variegated and severely asymmetrical, is a consequence of 
the fact that radiation at a point is the composite of the contributions from many directions, 
with sources at a large variety of temperatures and emissivities,
from both adjacent rays and distant sources at other angles.  
This smoothing effect was demonstrated definitively in 2D by Ott et al. 
(2008) and Brandt et al. (2011), using multi-dimensional/multi-angle transport, and by Dolence, 
Burrows, \& Zhang (2015), using flux-limited diffusion.\footnote{{However, Dolence, Burrows, \& Zhang (2015),
in particular, used multi-group flux-limited diffusion (MGFLD),
which is known to smooth out radiation fields at large distances
in the transparent regime.}}
For instance, the magnitude of the variation with angle in either the neutrino flux
or the specific neutrino heating rate from pole to equator of a 2D simulation
not employing the ray-by-ray simplification
is typically less than $\sim$5\%.  {However, when employing the ray-by-ray
procedure, such variations are typically $\sim$20\% (Sumiyoshi et al. 2015; 
Dolence, Burrows, \& Zhang 2015). Lund et al. (2012) and Tamborra et al. (2013)
perform 3D simulations using ray-by-ray+ and see variations of $\sim$50\%{}}
\footnote{A commonly used variant, ``ray-by-ray+,"
includes a matter velocity advection step of either the neutrino number
or energy density and the radial flux component. See Appendix for details.}.
When it is often stated that small effects at the $\sim$10\% level in the
equation of state or neutrino-matter couplings might make a qualitative difference 
in the computational outcome of collapse simulations, such a large difference 
between ray-by-ray and real multi-dimensional transport should give one pause.  This is particularly 
relevant for 2D simulations in which axial sloshing is so pronounced, and when such 
dipolar oscillations have themselves been invoked as potentially central to the mechanism
of explosion (Hanke et al. 2012).  It should be noted that {\it all} groups currently reporting
explosions by the neutrino-heating mechanism employ the ray-by-ray dimensional reduction simplification
(with the ``plus" extension, or otherwise).

The only self-consistent calculations that employed truly multi-dimensional transport were
those of Burrows et al. (2006,2007) and Ott et al. (2008), using the VULCAN/2D code, and Dolence, 
Burrows, \& Zhang (2015), using the CASTRO code (Zhang et al. 2011,2013).  They did not 
obtain explosions in their 2D simulations.  Since most other self-consistent 
calculations that obtained explosions in 2D used the ray-by-ray scheme, while the VULCAN/2D
and CASTRO studies did not, one is tempted to suggest (as has Burrows 2013) that the ray-by-ray
dimensional reduction and simplification may in 2D be yielding qualitatively incorrect 
results. The ray-by-ray anomalies in the angular distribution of the 
radiation field and corresponding neutrino heating rates may be reinforcing the axial sloshing motions 
and in 2D pushing the shock into explosion.\footnote{The general absence in 3D of either an axial effect or the 
pronounced sloshing seen in 2D may be rendering 3D simulations performed with the 
ray-by-ray approach less problematic.} 

Hence, the current explosions in 2D may in part be numerical artifacts. Unfortunately, 
VULCAN/2D (Burrows et al. 2007) and CASTRO (Zhang et al. 2011,2013) employed 
Cartesian grids in the central regions, and, hence, were not suitable to test 
in a self-consistent fashion the ray-by-ray approach to multi-dimensional collapse simulations.  
Surprisingly, the accuracy of the ray-by-ray+ scheme has never before been
tested, not by us nor by those who use it, and such a self-consistent head-to-head
comparison in the time-dependent context is the major motivation for this paper.
One of the motivations for the development of our new code F{\sc{ornax}} (Dolence, Burrows, \& Skinner 
2016) was to be able to compare directly, and with the same code, results using more realistic 
multi-dimensional transport against those using the ray-by-ray simplification.
F{\sc{ornax}}, with its spherical geometry and multi-dimensional capability, can do this.

In section \S\ref{code}, we briefly describe our F{\sc{ornax}} code and the method we use to implement 
the ray-by-ray+ simplification.  A more complete description of F{\sc{ornax}} can be found 
in Dolence, Burrows, \& Skinner (2016).  Section \S\ref{results} contains the results of our comparison of 
the ray-by-ray and multi-dimensional radiation/hydro simulations in 2D, with a particular focus
on the different hydrodynamical responses, shock positions and motions, and overall outcomes.
We conclude in section \S\ref{conclusions} with a discussion of what we have found concerning 
the limitations of the ray-by-ray approach.  {We emphasize that this paper is not meant to be 
a full supernova mechanism study.  Rather, in this paper we focus solely on the viability and 
accuracy of the ray-by-ray+ transport methodology and leave more comprehensive 
studies using F{\sc{ornax}} of the supernova phenomenon per se to later work.}

\section{Methodology}
\label{code}

Since the speed of sound is close to the speed of light 
for most of the evolution of a proto-neutron star core, in 
the core-collapse supernova problem we can integrate the transport 
operator {\it explicitly}, in a manner analogous to that employed by 
Skinner \& Ostriker (2013), but without the need for an artificially reduced
speed of light in the post-bounce phase. {This has also been noticed by 
Just et al. (2015) and O'Connor \& Couch (2015) and obviates the need to 
do implicit global iterations, speeding up the calculations for a given core count by
a {\it factor} of $\sim$5$-$10.}\footnote{{Much of that speed up 
is due simply to avoiding the many iterations to convergence of the spatial 
derivative terms required in such implicit methods, and the associated global 
communication overhead.}}

Our implementation of this is F{\sc{ornax}}, 
purpose-built by our team over the last two years.  
{The code solves the
equations of self-gravitating, compressible hydrodynamics with an
arbitrary (often tabulated) equation of state, coupled to the
multigroup two-moment equations for neutrino transport. We solve the
equations on a fixed Eulerian grid, where the geometry and
coordinates are handled in a generally-covariant way using
a spatial metric and associated Christoffel symbols.  Thus,
F{\sc{ornax}} can easily incorporate a variety of geometries
and coordinate choices, with the restriction that the metric
and coordinate transformations remain orthogonal.  For this work,
we have chosen a spherical geometry and adopt a radial
coordinate $x_1$ such that $r=x_t \sinh (x_1/x_t)$, where $x_t$
is solved to give a specified central resolution (0.5 km) and 
outer radius ($10^4$ km).  Roughly, $x_t$ is associated with
the radius at which a uniform grid in $x_1$ transitions from
a uniform grid in $r$ to a uniform grid in $\log r$.  In multiple
dimensions, a spherical grid that extends down to $r=0$ can be
restrictive as the Courant-limited timestep can become extremely
small due to the convergence of radial grid lines.  While others have
sometimes simply evolved the inner region in 1-D spherical symmetry
to overcome this problem, we instead enhance our otherwise logically
Cartesian mesh with a simple form of static mesh derefinement.
Regardless of the resolution specified in the angular direction,
the derefined mesh has only two zones in each angular direction
for those zones that reach $r=0$.  Subsequently, the number of
zones in the angular direction is doubled for every doubling of
radius until a specified number of cells is reached.  The outcome
of this process is 1) a mesh where the cell aspect ratios are
never extreme, 2) a true representation of the multi-D dynamics throughout
the whole domain, and 3) a drastically improved timestep limit.
\footnote{Since cell aspect ratios are kept within a factor of two
of unity, the explicit CFL timestep is kept within a factor of two
of that of the innermost radial zone.} This type of mesh is sometimes referred to as dendritic, having
nodes on zone edges in some cells instead of exclusively at corners.
On our Eulerian mesh, we employ a finite-volume discretization and
evolve our equations with a directionally unsplit higher-order
Godunov-type scheme with Shu \& Osher's optimal second-order Runge-Kutta
time stepping (Shu \& Osher 1988).  Hydrodynamic and compositional
variables are conservatively reconstructed using a novel ``geometry-aware''
parabolic profile to compute edge states that define a Riemann problem 
on each face.  With these edge states, the intercell fluxes of the 
conserved hydrodynamic variables are computed with the HLLC Riemann solver of 
Toro et al. (1994).} 

The multi-group two-moment equations for neutrino
transport are formulated in the comoving frame and include 
all terms to $O(v/c)$. The moment hierarchy is closed with the ``M1'' model
(Vaytet et al. 2011), though Fornax can easily accommodate other closure choices.
After core bounce, the fastest hydrodynamic signal speeds in the
core-collapse supernova problem are within a factor of a few of
the speed of light, so a time-explicit integration of the
non-stiff transport terms is not only simpler and generally more
accurate, it is also faster than globally coupled time-implicit
transport solvers that are typically employed in radiation hydrodynamics
methods.  {Radiation quantities are reconstructed with linear profiles
and the resulting edge states are used to calculate fluxes via the
HLLE solver, similar to Vaytet et al. (2011).  As in O'Connor \& Ott (2013),
the HLLE fluxes are corrected to reduce numerical diffusion in the
non-hyperbolic regime.  The terms that transfer momentum and energy
between the radiation and the gas are operator-split and treated implicitly.
These terms are purely local to each cell and, therefore, do not introduce
any global coupling.  While in the free-streaming regime of transport
the solution is agnostic to the ordering of the operator-split updates,
the tightly coupled diffusive regime is sensitive to this choice. In F{\sc{ornax}},
the order of the operator-split updates are chosen to ensure the correct
asymptotic behavior in the diffusive regime.} {In an earlier draft of the 
paper, we performed our calculations without general-relativistic (GR) corrections,
but for the current version we have implemented and employed the effects of GR
in the manner of Rampp \& Janka (2002) and Marek et al. (2006).}

For the neutrino-matter coupling, the interaction rates of Burrows,
Reddy, \& Thompson (2006) are used. These incorporate weak-magnetism and
recoil corrections to the neutrino-nucleon interactions, screening and
ion-ion correlation corrections for the neutral-current scattering off
nuclei, and inelastic scattering of neutrinos off electrons. The latter
is handled using the approximate scheme of Thompson, Burrows, \& Pinto (2003),
which treats various terms of the scattering matrix either explicitly or
implicitly so that inelastic scattering can be written as a sum of effective
source and sink terms in and out of an energy bin. This approach is stable,
avoids the requirement of energy bin couplings, and scales linearly with
energy group number. However, the associated cross sections are low and
final-state blocking (and, hence, suppression) is significant (particularly
after bounce). Therefore, for this study we ignore inelastic scattering.
We use the Lattimer-Swesty nuclear equation of state with K = 220 MeV for
all simulations (Lattimer \& Swesty 1991).

For this study comparing 2D simulation results using ray-by-ray and actual  
multi-dimensional transport, we employ F{\sc{ornax}} with 40 energy 
($\varepsilon_{\nu}$) groups for each of the $\nu_e$, $\bar{\nu}_{e}$, and $\nu_{\mu}$
species, where $\nu_{\mu}$ represents $\nu_{\mu}$, $\bar{\nu}_{\mu}$, $\nu_{\tau}$, and $\bar{\nu}_{\tau}$ types 
collectively.  For the $\nu_e$ types, $\varepsilon_{\nu}$ varies logarithmically from 1 to 300 MeV, while it varies
from 1 to 100 MeV for the other species.  For the pair processes, we assume detailed
balance, but ignore the blocking of two final-state neutrino species. 
Spherical coordinates are used.  The radial coordinate, $r$,
runs from 0 to 10,000 kilometers (km) in 608 zones. 
The radial grid smoothly transitions from uniform spacing with $\Delta r=0.5$ km 
in the interior to logarithmic spacing, with a characteristic transition
radius of $\sim$100 km.  The angular grid spacing varies smoothly from 
$\approx 0.95^\circ$ at the poles to $\approx 0.65^\circ$ at the equator in 256 zones,
covering the full $180^\circ$.  We discuss the issue of resolution in \S\ref{resol}.
The 12-, 15-, 20- and 25-M$_{\odot}$ massive-star progenitors we use as initial models are from Woosley 
\& Heger (2007).  The 2D simulations were performed on the Cray machines Blue Waters and Cori
and take approximately two days each (using 1024 cores) to complete.

To perform ray-by-ray+ calculations, we set all transverse neutrino radiation fluxes 
equal to zero, effectively updating only the radial components of the flux in each energy group and for each 
neutrino species.  However, we implement the ``ray-by-ray+" variant by retaining the velocity advective
terms in the zeroth-moment (energy) and first-moment (momentum) radiation equations (Buras et al. 
2006).  The Appendix highlights these various terms as they appear in our formulation.  This also
requires retaining the second- and third-moment pressure and heat-flux tensor 
components in the geometric source terms associated with the Christoffel 
terms that arise in spherical coordinates (in particular to retain transverse pressure 
gradients). We note that other groups using other transport implementations (e.g., 
multi-group flux-limited diffusion [MGFLD]; spherical Boltzmann; IDSA) will be 
constrained to use their own closures for these source and advective terms 
(if implementing the ``plus" variant), but that the ``transverse flux equals 
zero" step likely has the greatest consequences and is the central assumption
of ray-by-ray.  

{The original version of
F{\sc{ornax}} was built with flexibility in mind, and geometrical quantities like
face areas and cell volumes were numerically computed via quadrature. Though
these integrals were computed extremely precisely, a very low-level of
symmetry breaking was introduced. This was particularly important
in the inner region where we employ static mesh refinement (the dendritic
grid).  The new version now eliminates this problem.  For the calculations in this paper,
we impose initial perturbations \`a la M\"uller and Janka (2015) with n=5, l=2,
and v$_{max}$ = 1000 km s$^{-1}$, and do not allow the grid and computational noise
to dictate initial asphericities.}

{One remaining source of unphysical symmetry breaking is the tabular equation of state.
For the calculations presented in this work, we employed a Newton-Raphson
iteration to compute thermodynamic quantities as a function of density,
specific internal energy, and electron fraction from a table that provides
values as a function of density, temperature, and electron fraction. Part
of this process relies on an initial guess for the temperature and the
iteration proceeds until a relative tolerance of 10$^{-8}$ is reached.  Given
two different initial guesses, the iteration will not result in identical
thermodynamic outputs even for identical densities, internal energies, and
electron fractions.  These differences are small, $\sim$$\epsilon T
\partial q/\partial T$, where epsilon is the relative tolerance and q is
the quantity in question.  We note that for multi-dimensional models
symmetry breaking is required.  Most authors rely on grid level noise
of some sort to achieve this. Importantly, whatever the source of noise
in our calculations, it is exactly the same for both ray-by-ray+ and
non-ray-by-ray+ multi-D runs.}

\section{Results}
\label{results}

\begin{figure}
   \plottwo{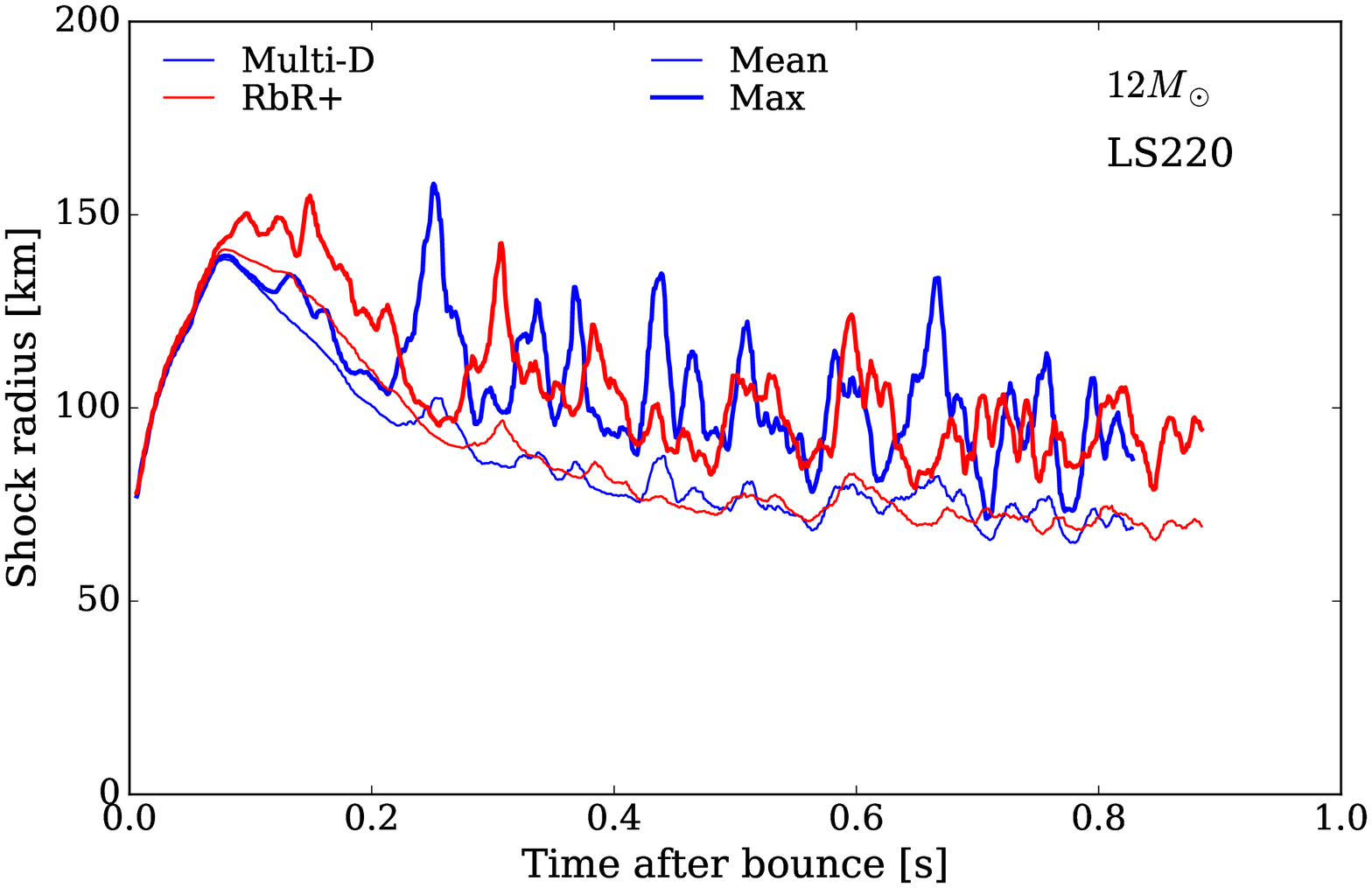}{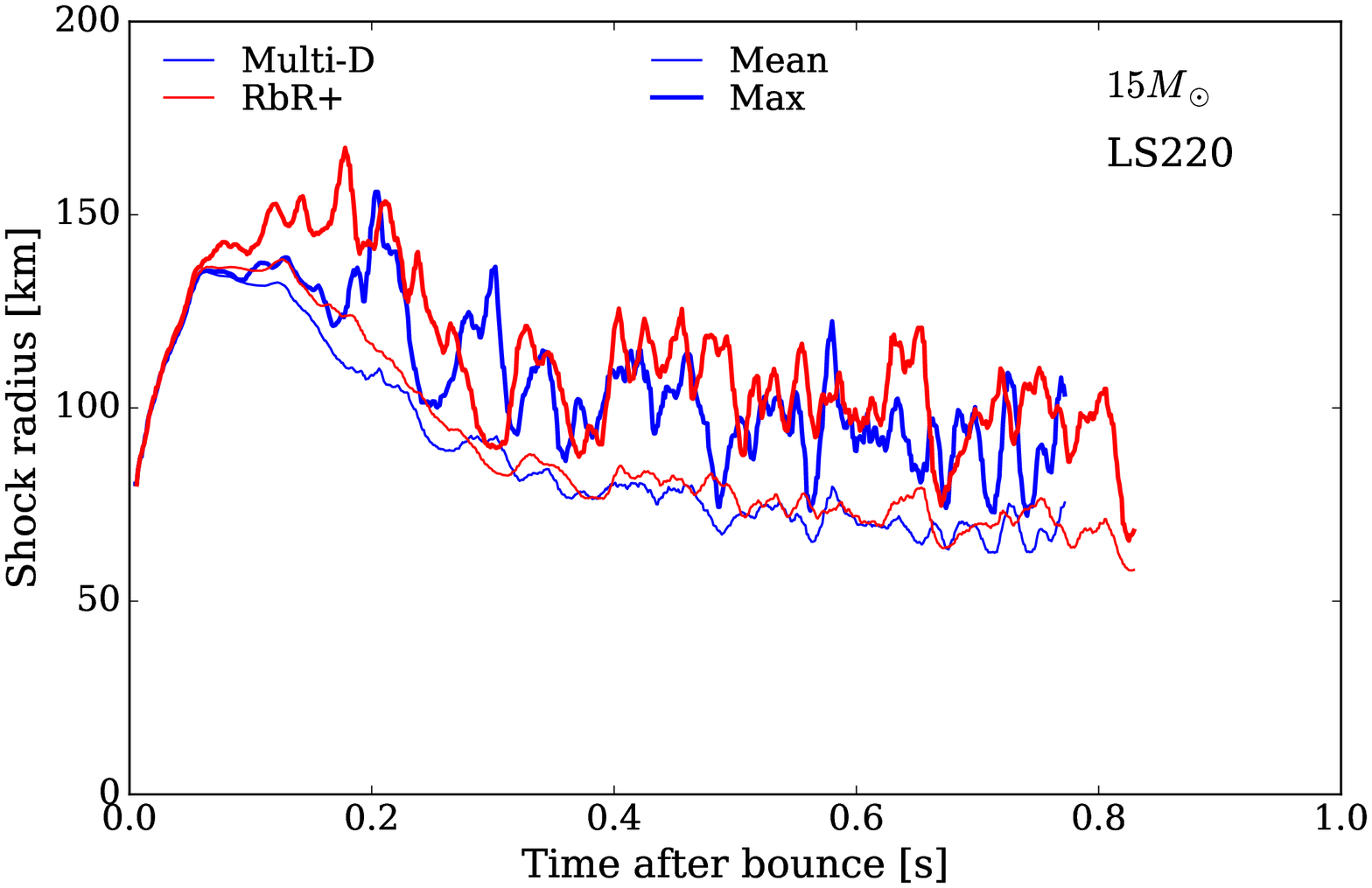}
\caption{{\bf Left:} The evolution of the average (thin line) and maximum (thick line) shock radii
versus time since core bounce for the the 12-M$_{\odot}$ progenitor model of Woosley 
\& Heger (2007), both for the full transport model
using the two-moment closure multi-dimensional
transport scheme (blue) (Dolence et al. 2016) and the corresponding ray-by-ray+ 
model (red). Time is in seconds and the radii are in kilometers (km).  
{\bf Right:} The same as in the left-hand panel, but
employing the 15-M$_{\odot}$ progenitor model of Woosley    
\& Heger (2007). For both the left and the right panels, a 10-millisecond boxcar
convolution was applied.
}
\label{ray_by_ray.12}
\end{figure}

\begin{figure}
  \plottwo{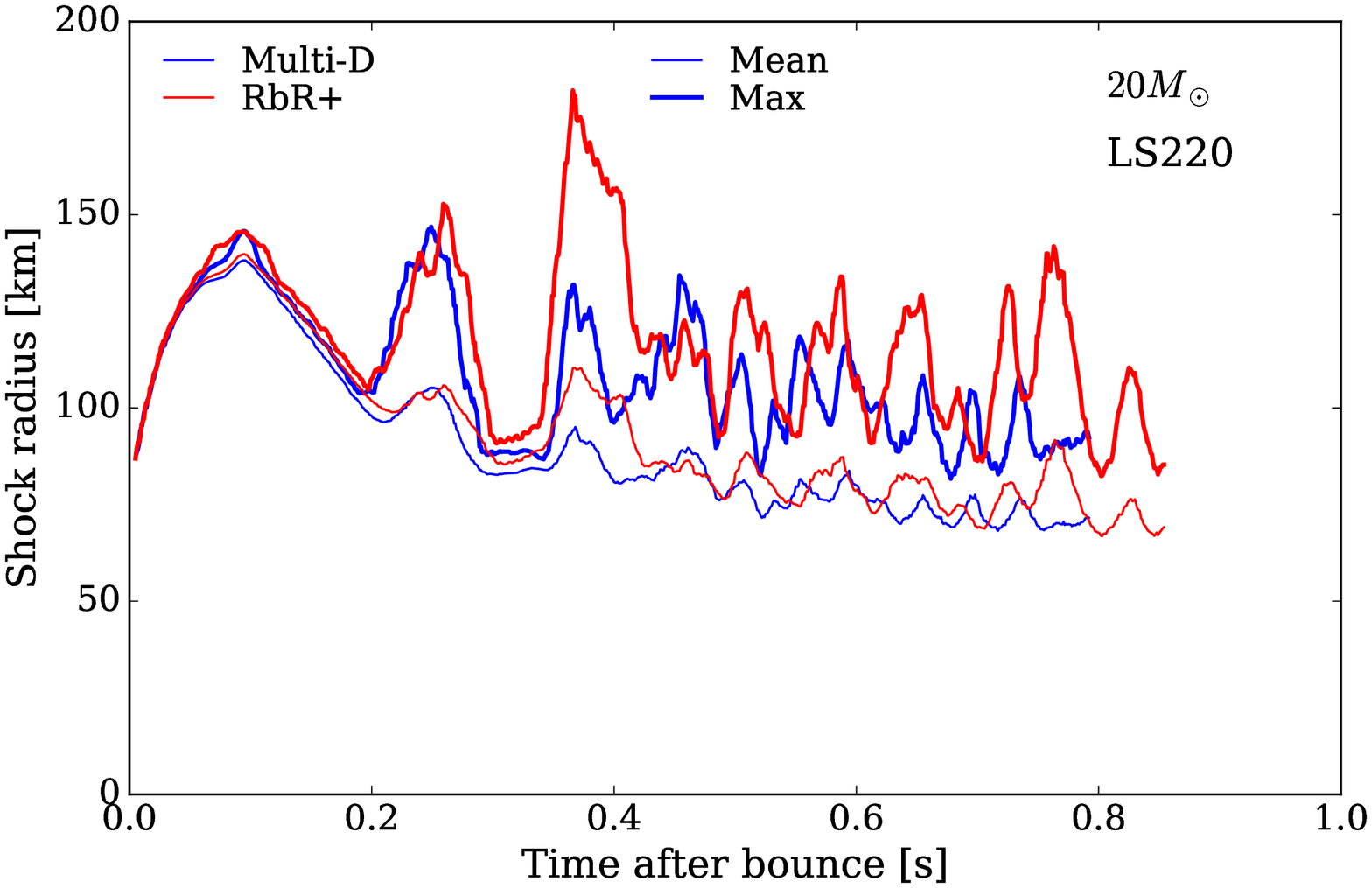}{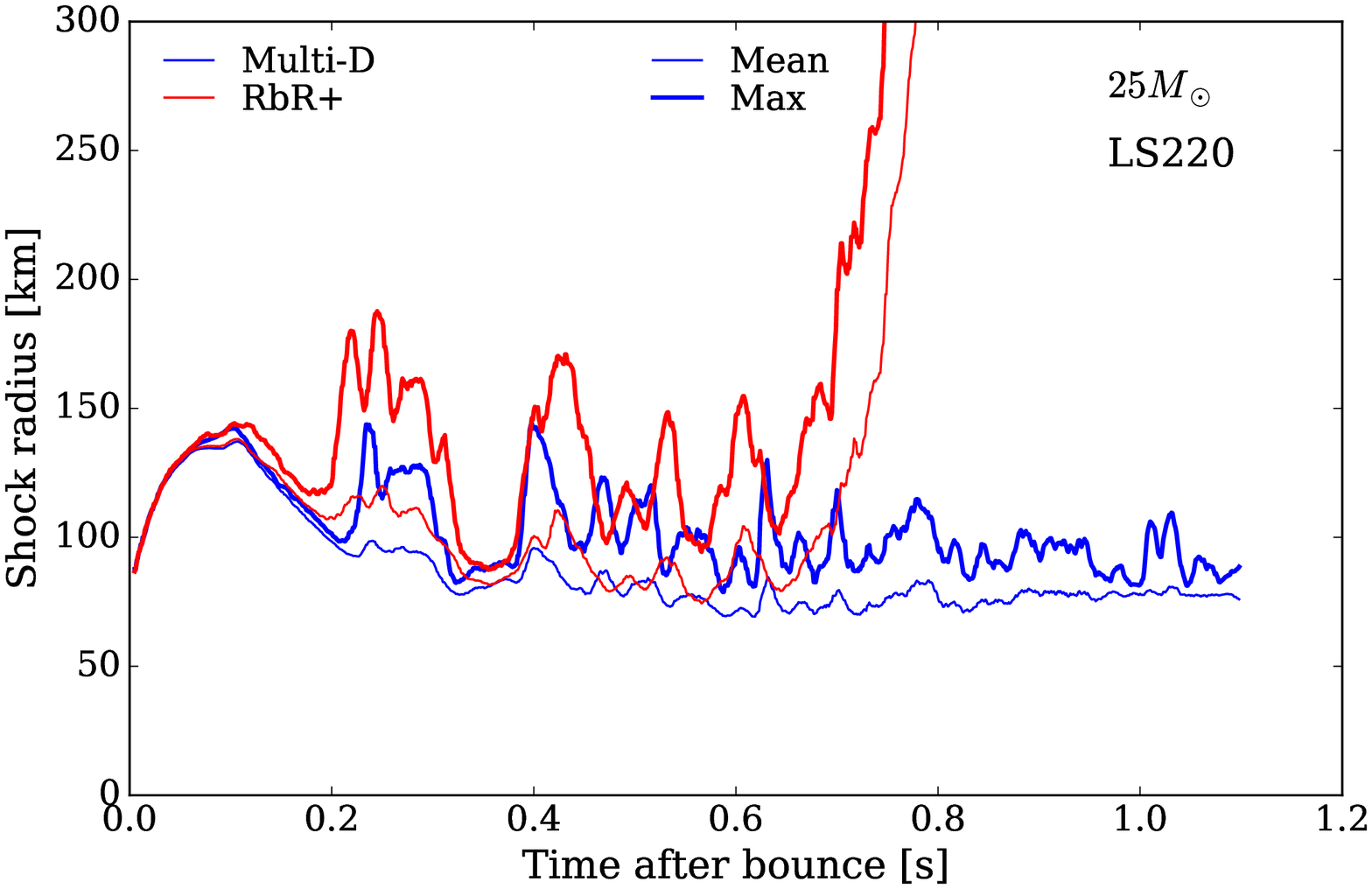}
\caption{Same as for Figure \ref{ray_by_ray.12}, but for the 20-M$_{\odot}$ and 
25-M$_{\odot}$ progenitor models of Woosley \& Heger (2007). Note that the ordinate scale
for the right-hand panel has been expanded to 300 kilometers.  While the ray-by-ray+ 
variant of the 25-M$_{\odot}$ model explodes, the corresponding multi-D model does not, 
even after 1.1 seconds of post-shock evolution. See text for a discussion.
}
\label{ray_by_ray.20}
\end{figure}

\begin{figure}
  \plotone{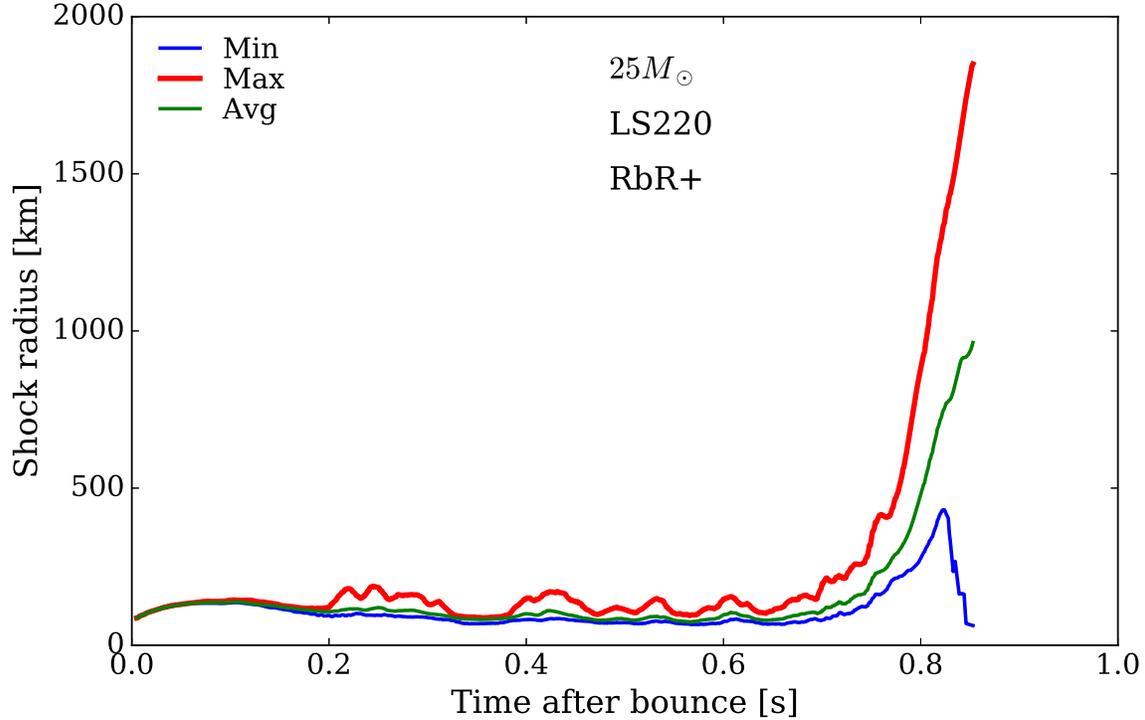}
\caption{The minimum, average, and maximum shock radii of the 25-M$_{\odot}$ 
progenitor model using ray-by-ray+ method that explodes at late times.  This is the same as 
depicted in the right panel of Figure \ref{ray_by_ray.20}, but on a much expanded vertical scale
in order  to better show the explosion.  The maxmium shock radius reaches $\sim$2000
kilometers by the end of the simulation, while the corresponding average radius
achieves $\sim$1000 kilometers, indicating the asymmetry of this 2D blast.
}
\label{explosion}
\end{figure}

In order to assess the differences between multi-D transport
and the ray-by-ray+ simplification when investigating the dynamics 
and outcomes of multi-D core collapse, we have simulated the evolution
of the collapse, bounce, and subsequent behavior of four non-rotating
progenitor models from Woosley \& Heger (2007).  These 12-, 15-, 20-, and 25-M$_{\odot}$
models were those employed by Bruenn et al. (2013,2014) in their recent 
explosion study, for which they used ray-by-ray+. Figures \ref{ray_by_ray.12}$-$\ref{ray_by_ray.20} depict
the evolution of the average (thin line) and maximum (thick line) shock radii ($R_{avg}$ and $R_{max}$) versus
time after bounce for the four models.  Superposed for easier comparison are
the multi-D transport F{\sc{ornax}} simulations (blue) and the corresponding ray-by-ray+ 
developments (red). First, we note that the post-shock region is chaotic and turbulent due to 
a combination of neutrino-driven convection and the standing-accretion-shock instability 
(SASI; Blondin et al. 2003). Hence, the detailed hydrodynamic behavior is stochastic and will 
depend upon resolution, the character of perturbations, and numerical implementation.  Nevertheless, 
Figures \ref{ray_by_ray.12}$-$\ref{ray_by_ray.20} show what are at times striking quantitative differences. 
{The maximum shock radii ($R_{max}$) for the low-accretion-rate 12-M$_{\odot}$ and 15-M$_{\odot}$ 
models are not much different between the multi-D transport and ray-by-ray+ realizations,
though the average excess for the ray-by-ray+ and full simulations in $R_{max}$ is 
$\sim$5 kilometers.  However, for the 20-M$_{\odot}$ and 25-M$_{\odot}$ progenitors 
$R_{max}$ is systematically higher for the ray-by-ray+ realizations 
at almost all post-bounce times, with excesses ranging $\sim$10$-$50 kilometers.  
In fact, Figures \ref{ray_by_ray.12}$-$\ref{ray_by_ray.20} demonstrate that the $R_{max}$ excess
for the ray-by-ray+ runs is an increasing function of progenitor mass, and $R_{max}$ 
for the full transport realizations is almost never larger
than for the ray-by-ray+ runs. Importantly, the 25-M$_{\odot}$ progenitor explodes 
at late times for the ray-by-ray+ simulation, while it does not for the multi-D transport
run, achieving an $R_{max}$ of $\sim$2000 kilometers by the end of the simulation.
Figure \ref{explosion} depicts the explosion on a more appropriately expanded scale.  
These shock radius comparisons for a range of 
progenitors collectively suggest that the use of the ray-by-ray+ simplification 
can result in quantitatively, and perhaps qualitatively, different evolution, particularly 
for the more massive progenitors.} The significance of the differences in the behavior of 
the shock radii in these 2D simulations lies in the fact that the 
ray-by-ray+ approach results in post-shock matter that is easier 
to unbind, because it resides at larger radii in the shallower
depths of the gravitational potential well. Hence, we speculate that ray-by-ray+ simulations
are more sensitive to the details of the numerical implementation, to the opacities, and to perturbations 
(perhaps due to the accretion of turbulent infalling matter).  In sum, they are more ``explodable,"
but artificially so.  {In fact, for our 25-M$_{\odot}$ progenitor we see an explosion
only for the ray-by-ray+ variant. In general, the fact that the ray-by-ray+ method results in
systematically larger excursions in the maximum shock radius suggests that
its use can make explosions in 2D artificially a bit more likely.}
   
\begin{figure}
  \centering
  \plottwo{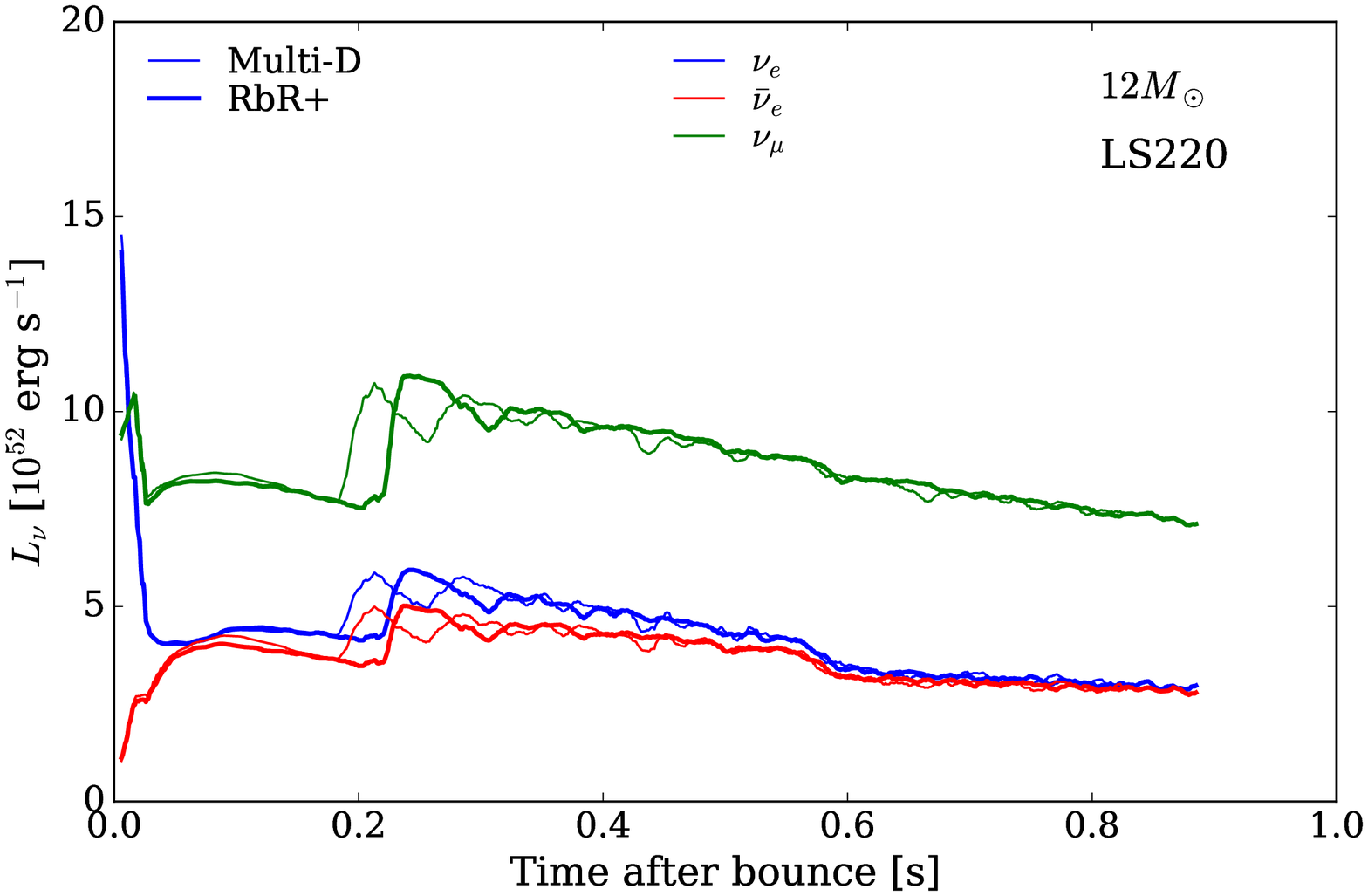}{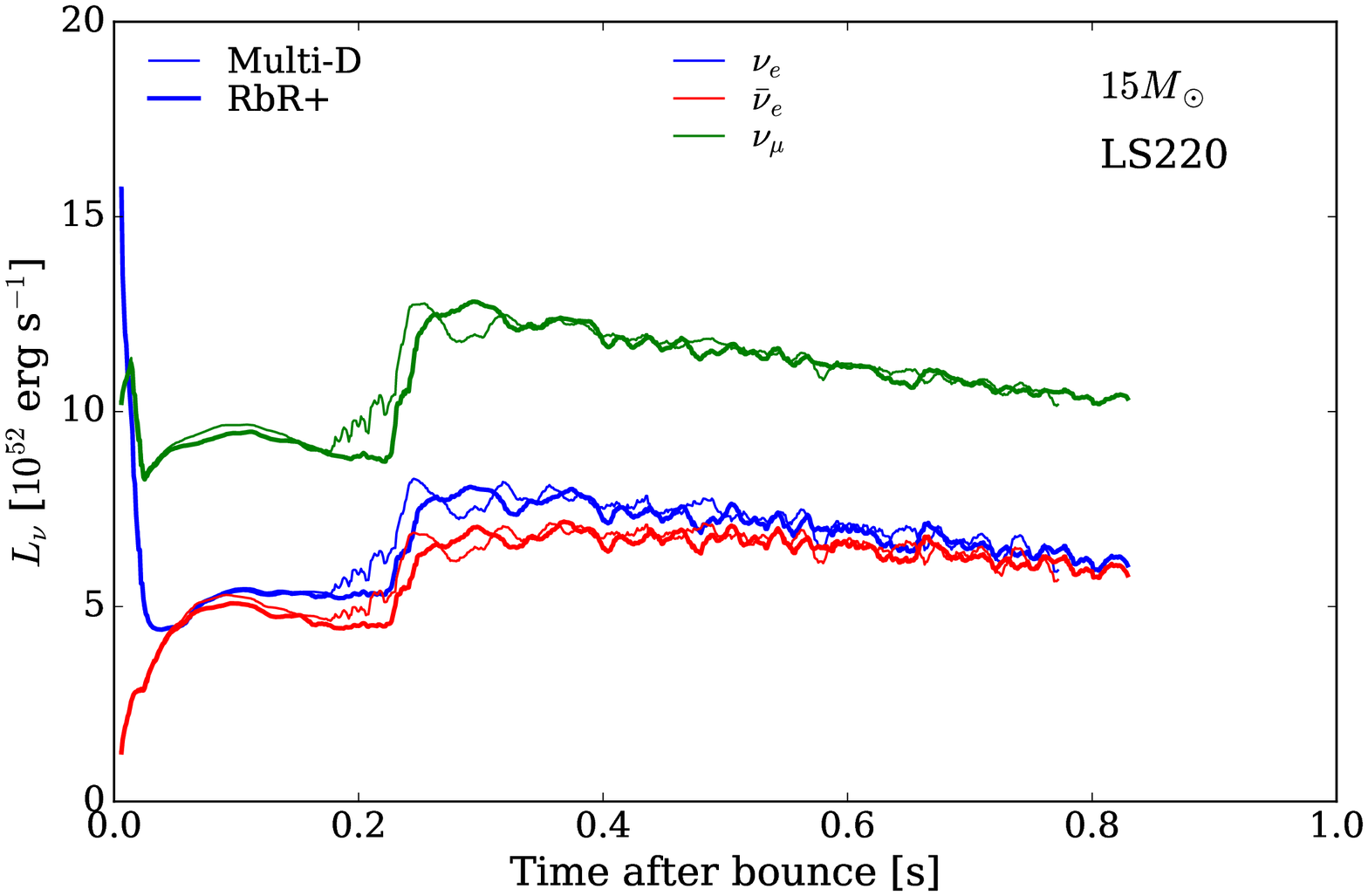}
\caption{The total solid-angle-integrated comoving-frame luminosities (in units of 10$^{52}$ ergs s$^{-1}$) 
at 100 kilometers of the $\nu_e$ (blue), $\bar{\nu}_{e}$ (red), and ``$\nu_{\mu}$" (green) 
neutrinos versus time after bounce (in seconds). The luminosity is smoothed over 10 milliseconds 
centered around this time.  The left-hand panel is for the 12-M$_{\odot}$ progenitor model,
and the right-hand panel is for the 15-M$_{\odot}$ progenitor model.  The thin solid curves are for the 
multi-D transport scheme, and the thick solid curves are for the ray-by-ray+ approach. 
{Note that the early bump in the $\nu_{\mu}$ luminosities results
from the fact we are calculating in the comoving frame and depicting the
result at 100 km.  The shock is formed deep and sweeps through
the inner radii at high speed. The bump we see is due to the change in
comoving-frame energy luminosity due to the abrupt change in the
velocity and the consequent Doppler shifts $-$ there would be no such 
bump if the calculations were in the lab frame, or the luminosity
were measured much further out.  We provide the luminosities at 100 km, 
not at 500 km or larger radii, since this radius is in the vicinity of 
the gain region.} 
}
\label{lum1}
\end{figure}

\begin{figure}
  \centering
  \plottwo{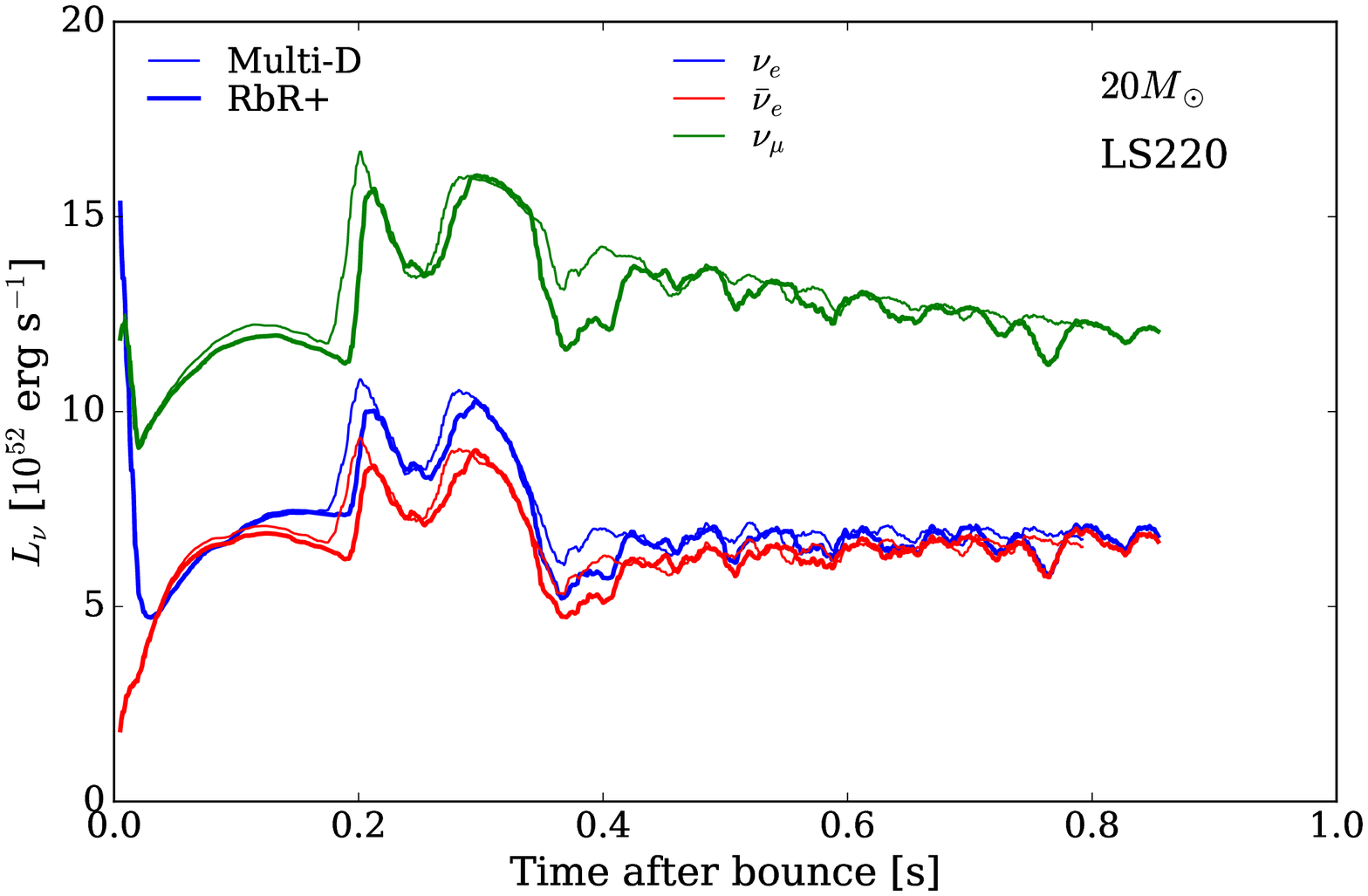}{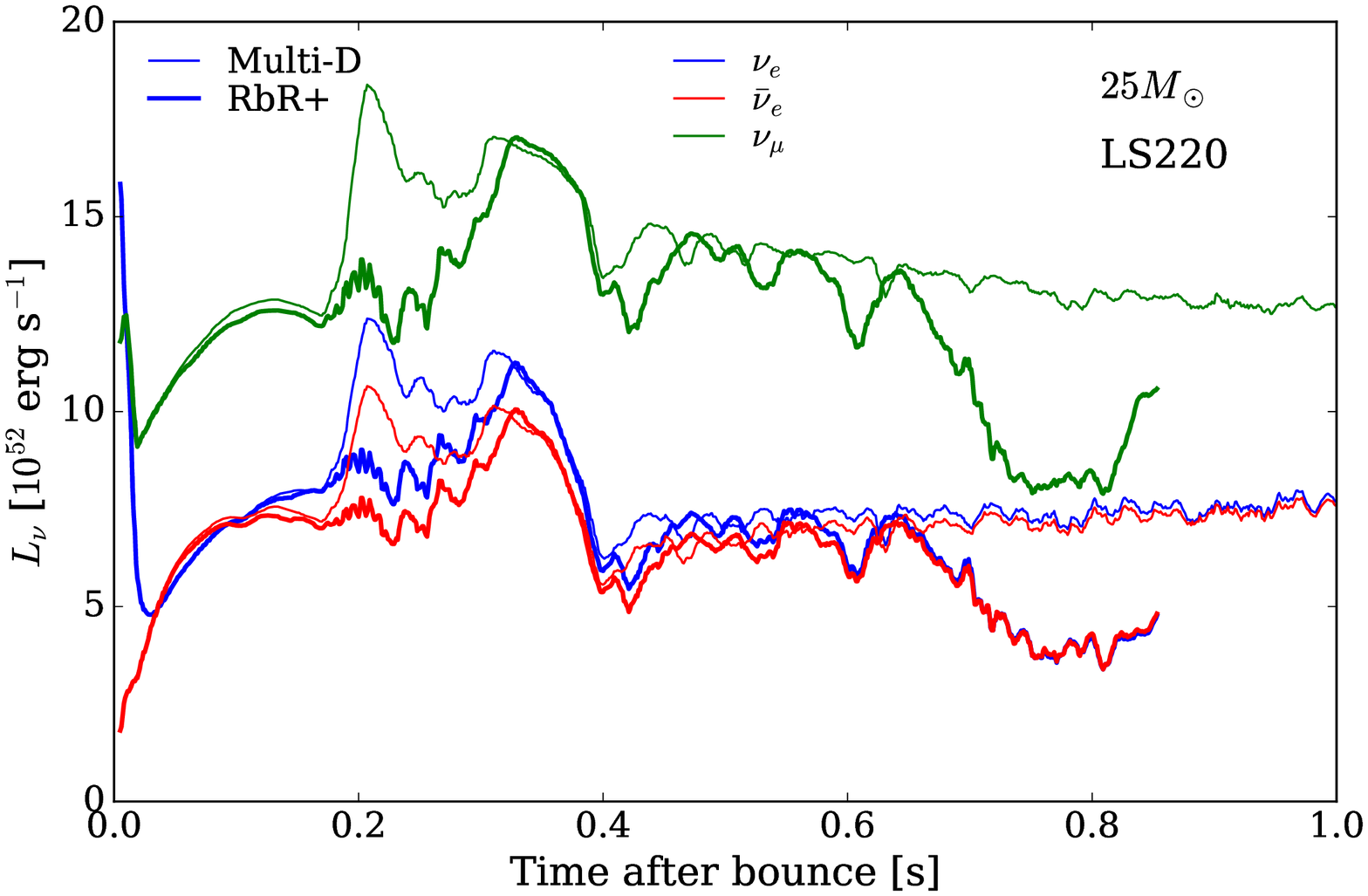}
\caption{Same as Figure \ref{lum1}, but for the 20-M$_{\odot}$ (left) and 25-M$_{\odot}$ (right) progenitor models. 
}
\label{lum2}
\end{figure}

The comoving-frame $\nu_e$ (blue), $\bar{\nu}_e$ (red), and ``$\nu_{\mu}$" (green)
luminosities at 100 km for all eight models, integrated over solid angle (and averaged over 10 milliseconds), 
are shown in Figures \ref{lum1} and \ref{lum2}.  The ray-by-ray+ values are rendered as thick solid lines.
As Figures \ref{ray_by_ray.12} through \ref{ray_by_ray.20} indicated, there are hydrodynamic
differences between the multi-D transport and ray-by-ray+ runs. {However, as 
Figures \ref{lum1} and \ref{lum2} indicate, the corresponding total luminosities 
are generally not much different, though for the 25-M$_{\odot}$ model the $\nu_{\mu}$ luminosity
can be slightly lower at later times. Since $\nu_{\mu}$s do not heat much, but mostly sap
the core of energy, this is conducive to explosion.  Also, for the 25-M$_{\odot}$ ray-by-ray+ model,
the total luminosities are lower for all species for about $\sim$100 milliseconds
around the time the Si/O interface is accreted at $\sim$0.3 seconds after bounce. The overall 
similarities in general, particularly for the 12-M$_{\odot}$, 15-M$_{\odot}$, and 20-M$_{\odot}$
progenitors, are likely because the quasi-dipolar oscillation of the post-shock material and shock 
position average out the accretion and internal flux components.}

\begin{figure}
  \centering
  \plottwo{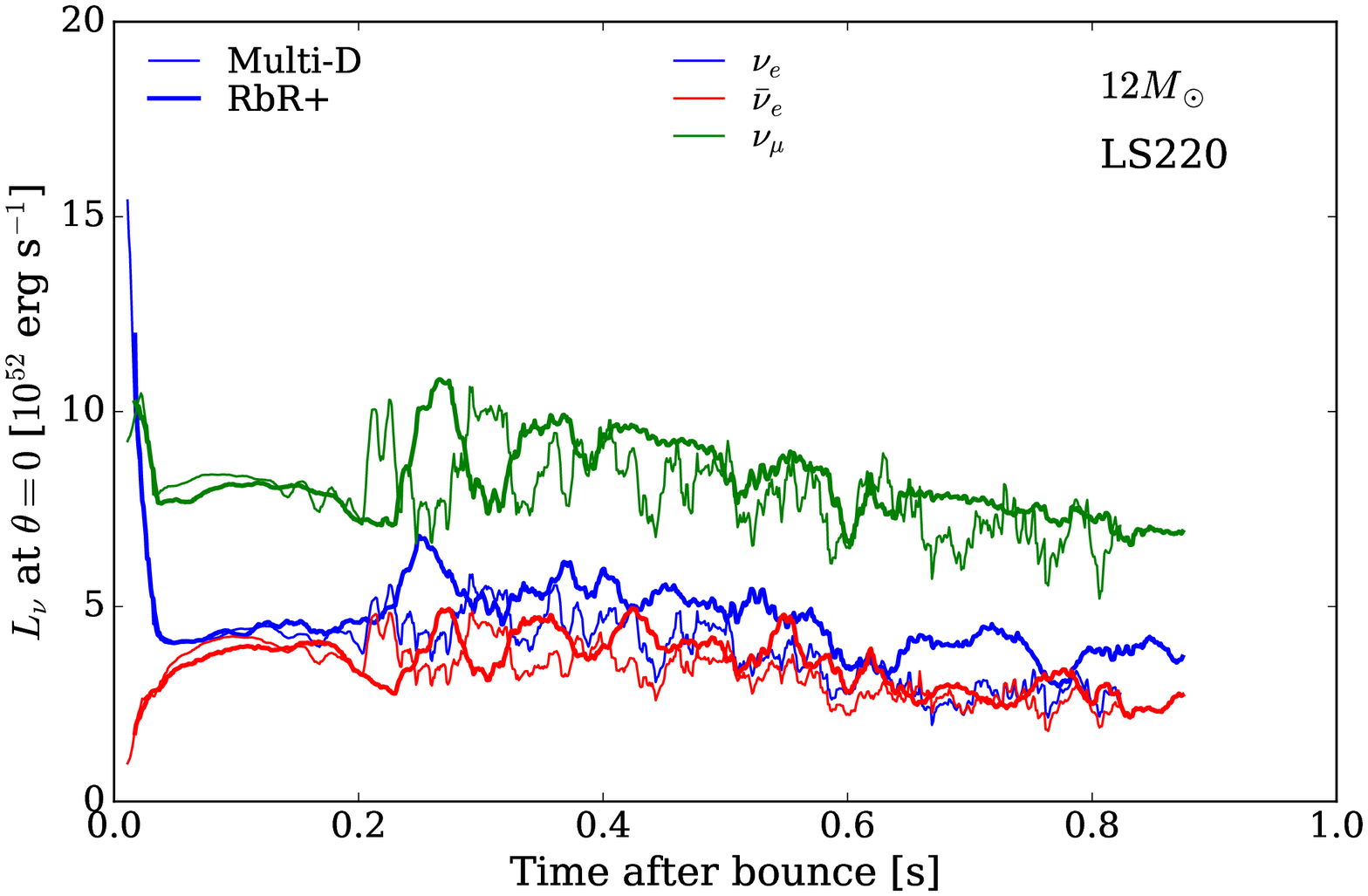}{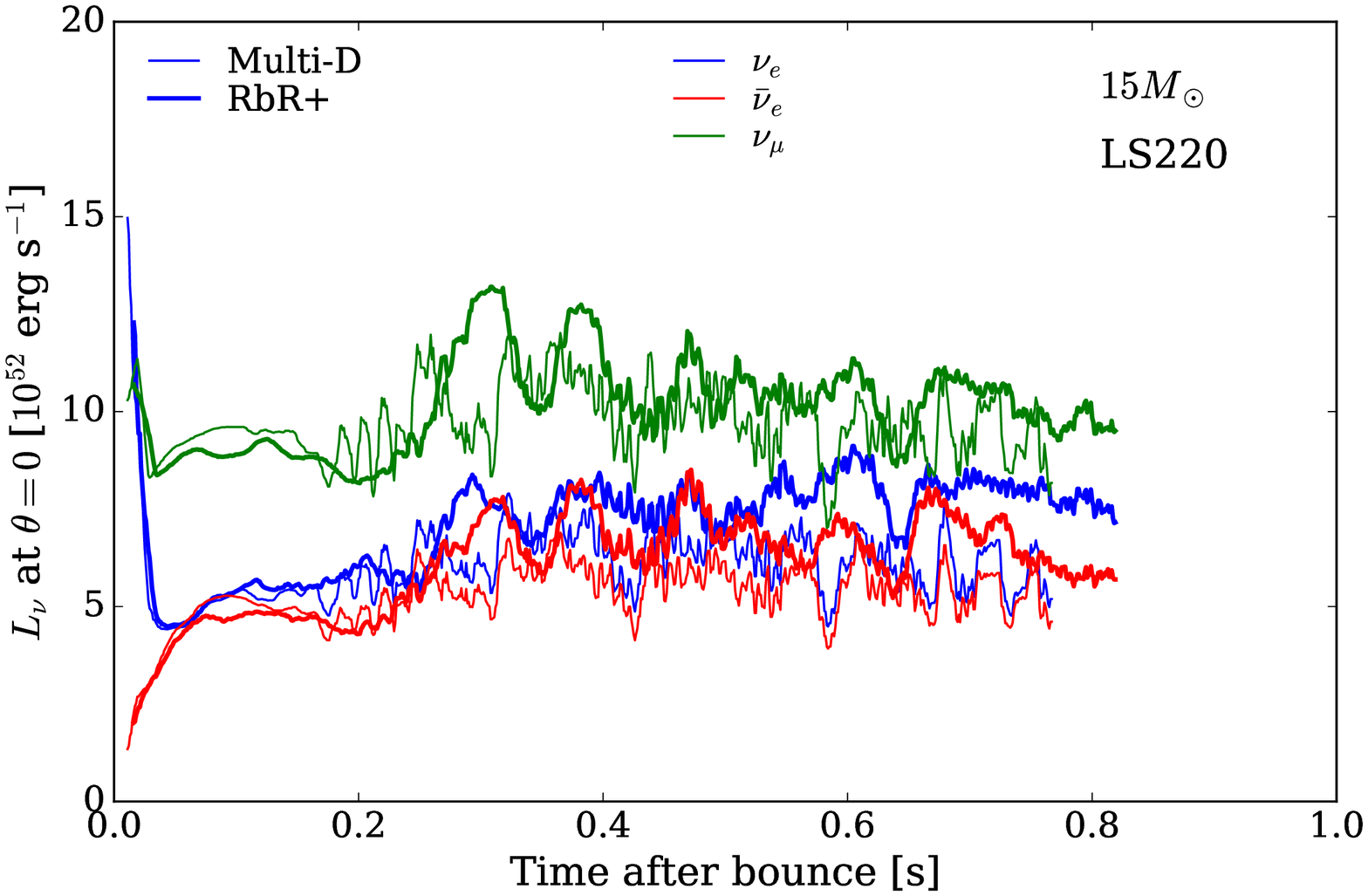}
\caption{Similar to the plots in Figures \ref{lum1} for the 12-M$_{\odot}$ and 15-M$_{\odot}$ progenitors, 
but for a polar luminosity constructed by averaging the radial flux within 20$^{\circ}$
of the pole and multiplying by $4\pi r^2$, where $r$ is 100 kilometers.  
Note that this pseudo-luminosity varies more significantly, 
and is demonstrably larger, for the ray-by-ray+ implementation, particularly at later times.  
A 20-millisecond boxcar convolution was applied to each graph.
}
\label{pole1}
\end{figure}

\begin{figure}
  \centering
  \plottwo{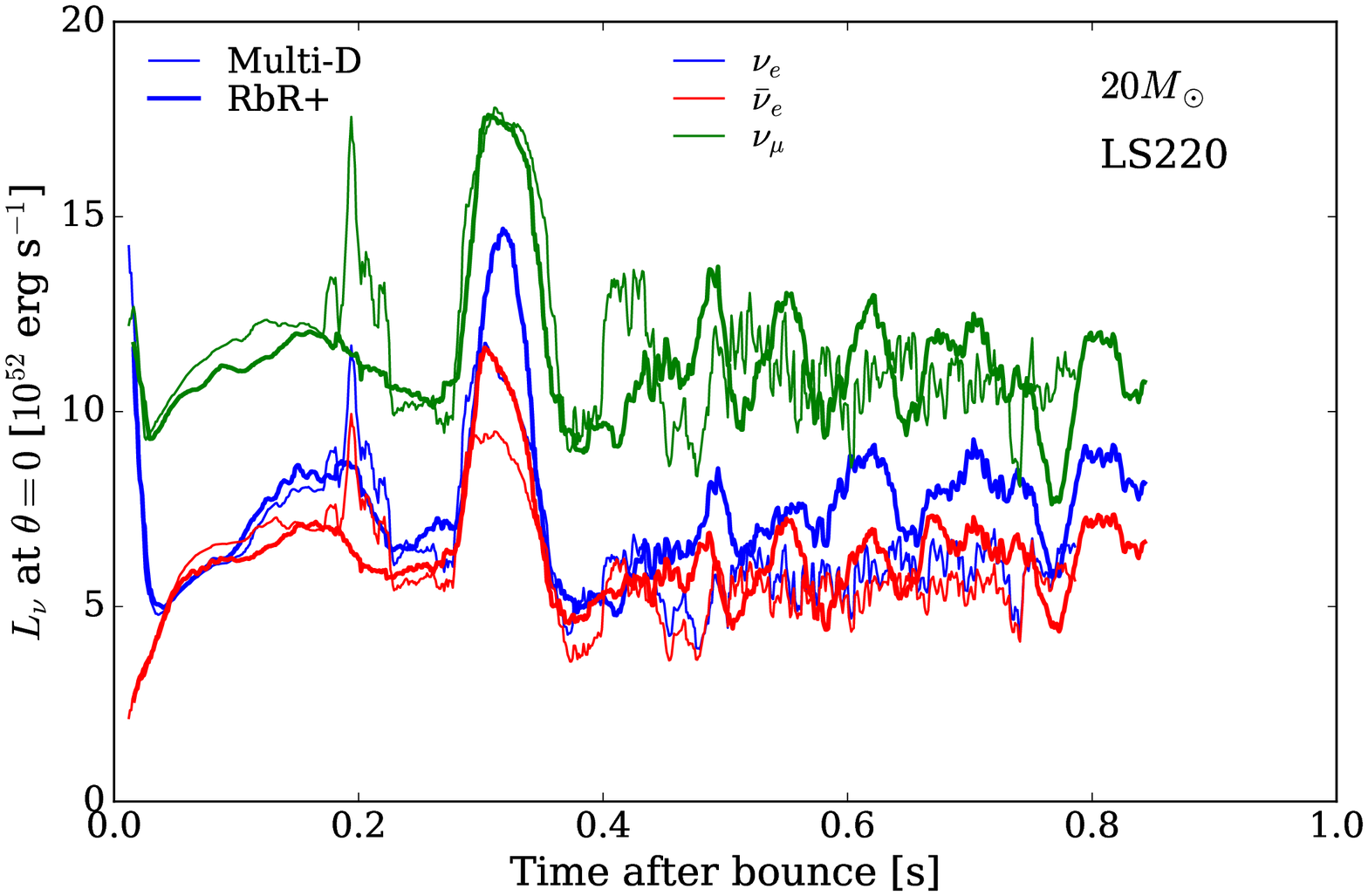}{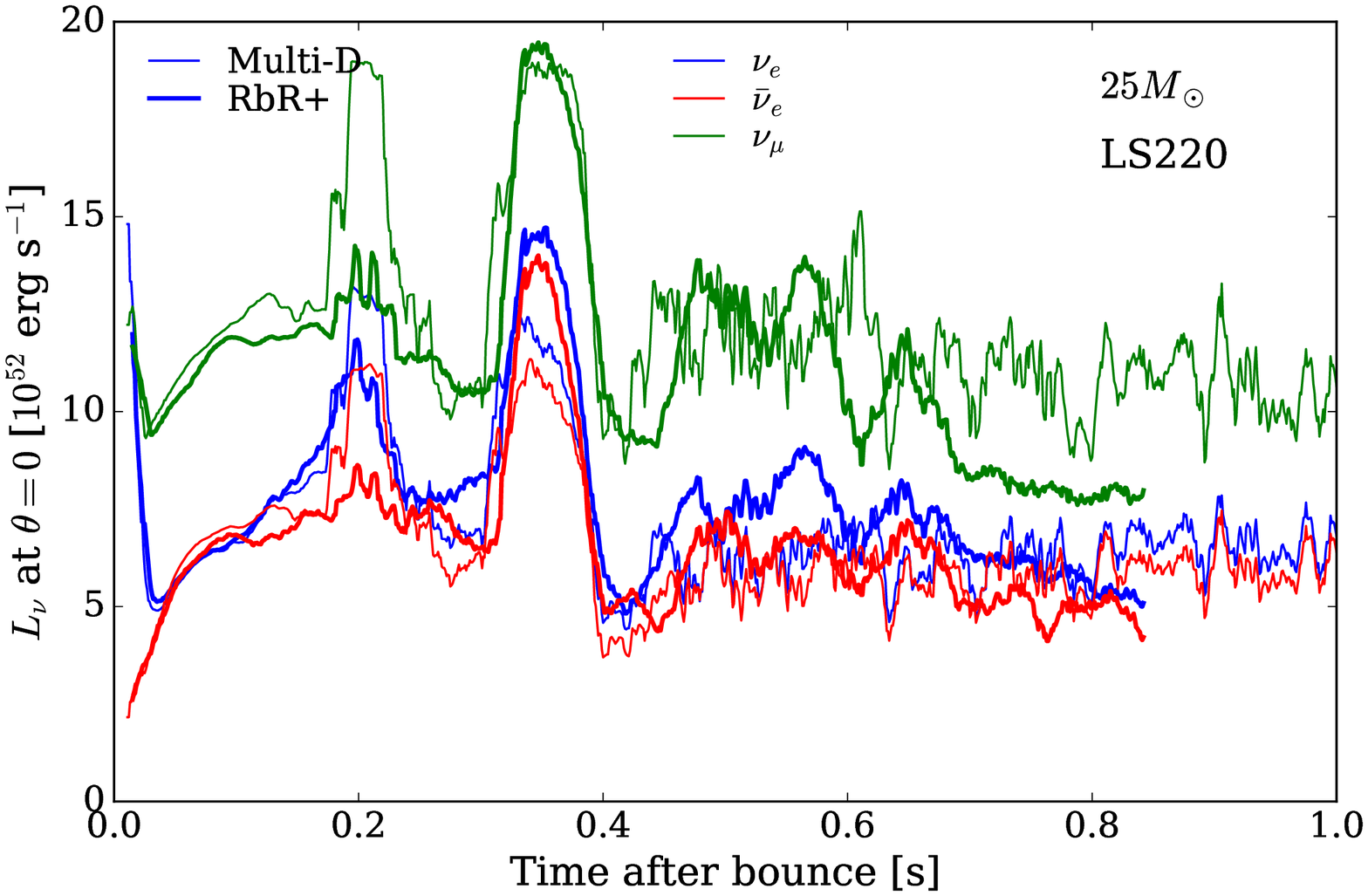}
\caption{Similar to the plots in Figures \ref{pole1}, but for the 20-M$_{\odot}$ and 25-M$_{\odot}$ progenitors. 
As in Figure \ref{pole1}, this pseudo-luminosity varies more significantly, and is demonstrably 
larger, for the ray-by-ray+ models.  
}
\label{pole2}
\end{figure}

\begin{figure}
  \centering
  \plottwo{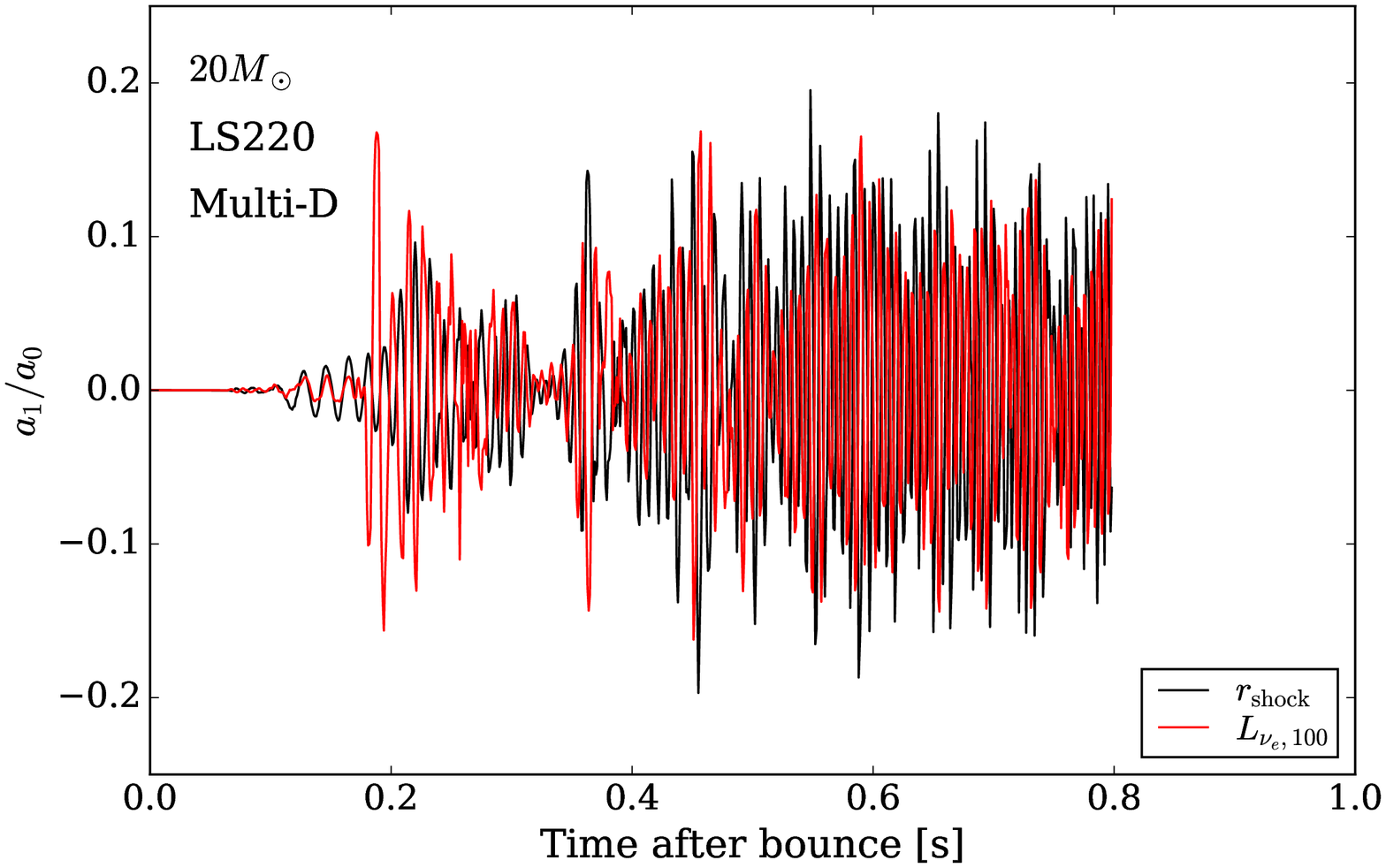}{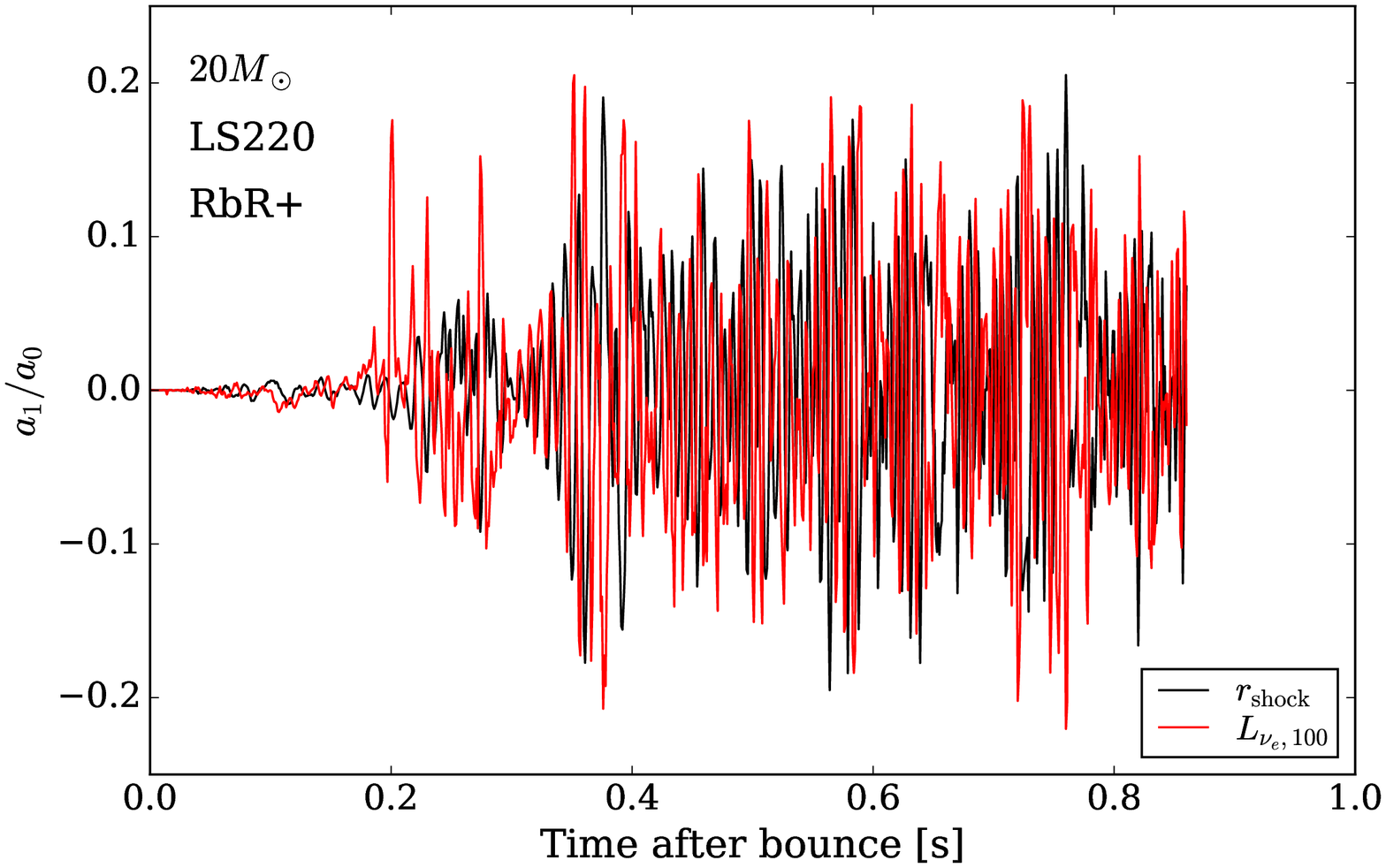}
\caption{The normalized angular dipole term (the dipole divided by the monopole) of the shock 
radius (black) and the energy-integrated $\nu_e$ luminosity (red) versus time after bounce (in seconds) 
for the 20-M$_{\odot}$ progenitor.  The left panel is for the full multi-D M1 simulation, and the right 
panel is for the associated ray-by-ray+ simulation.  These two dipole metrics track on another
rather well, for both the full transport and the ray-by-ray+ models. Note that significant
oscillation does not appear before about $\sim$150$-$200 milliseconds after bounce, reflecting 
the slow growth to non-linearity for the perturbations imposed. 
}
\label{dipole3}
\end{figure}

This is not as much the case
for the fluxes along the symmetry axes at $\theta = 0$ and $\theta = 180$.  If we calculate 
the average energy flux in a 20$^{\circ}$ wedge at the northern pole (defined as 
the $\theta = 0^{\circ}$ axis), and multiply by $4\pi r^2$ to obtain a pseudo-luminosity,
we see marked differences between the multi-D transport and the ray-by-ray+ numbers at later times.
{As Figures \ref{pole1} and \ref{pole2} indicate, the ``polar" $\nu_e$ and $\bar{\nu}_e$ 
luminosities evolve on average to be larger using the ray-by-ray+ approach $-$ 
as much as $\sim$30\% larger. The $\nu_e$ and $\bar{\nu}_e$ neutrino fluxes 
are most strongly coupled to matter by charged-current
absorption on liberated nucleons.} As has been pointed out previously 
(Burrows 2013; Dolence, Burrows, \& Zhang 2015; Sumiyoshi et al. 2015),
the ray-by-ray+ scheme makes the radiation field track the matter field too closely,
exaggerating its angular variation. True transport results in contributions 
to the neutrino energy density at a given point from the entire angular and radial domain, thereby
smoothing it out.  Aside from the modest effect of the advective terms, the ray-by-ray+ approach restricts
the influence at a point to the characteristics of the thermal and matter profiles 
($T$, $\rho$, and the electron fraction, $Y_e$) along only the associated radial ray.  This makes the radiation field 
unphysically sensitive to local and radial matter conditions, and, as we see, can
alter the matter motions.  Moreover, the variations
in the polar luminosities seen in Figures \ref{pole1} and \ref{pole2} are phased with the
shock oscillations, thereby leading to excess heating near when the matter is less bound.
Figure \ref{dipole3} for the 20-M$_{\odot}$ model comparison demonstrates this most dramatically.  
Here, we depict the normalized dipole component of both the shock radius and the $\nu_e$ luminosity for both
the multi-D transport and ray-by-ray+ simulations.  {We see on both panels 
that the amplitudes for both quantities are roughly in phase with one another.
However, the decrease seen in Figure \ref{pole2} by $\sim$30\% in the $\nu_{\mu}$ luminosity along
the poles for the ray-by-ray+ run of the 25-M$_{\odot}$ progenitor at late times 
may be suggestive, since $\nu_{\mu}$ neutrino emission is largely detrimental to
explosion; such a decrease may have aided this model to explode.}

\begin{figure}
  \centering
  \plottwo{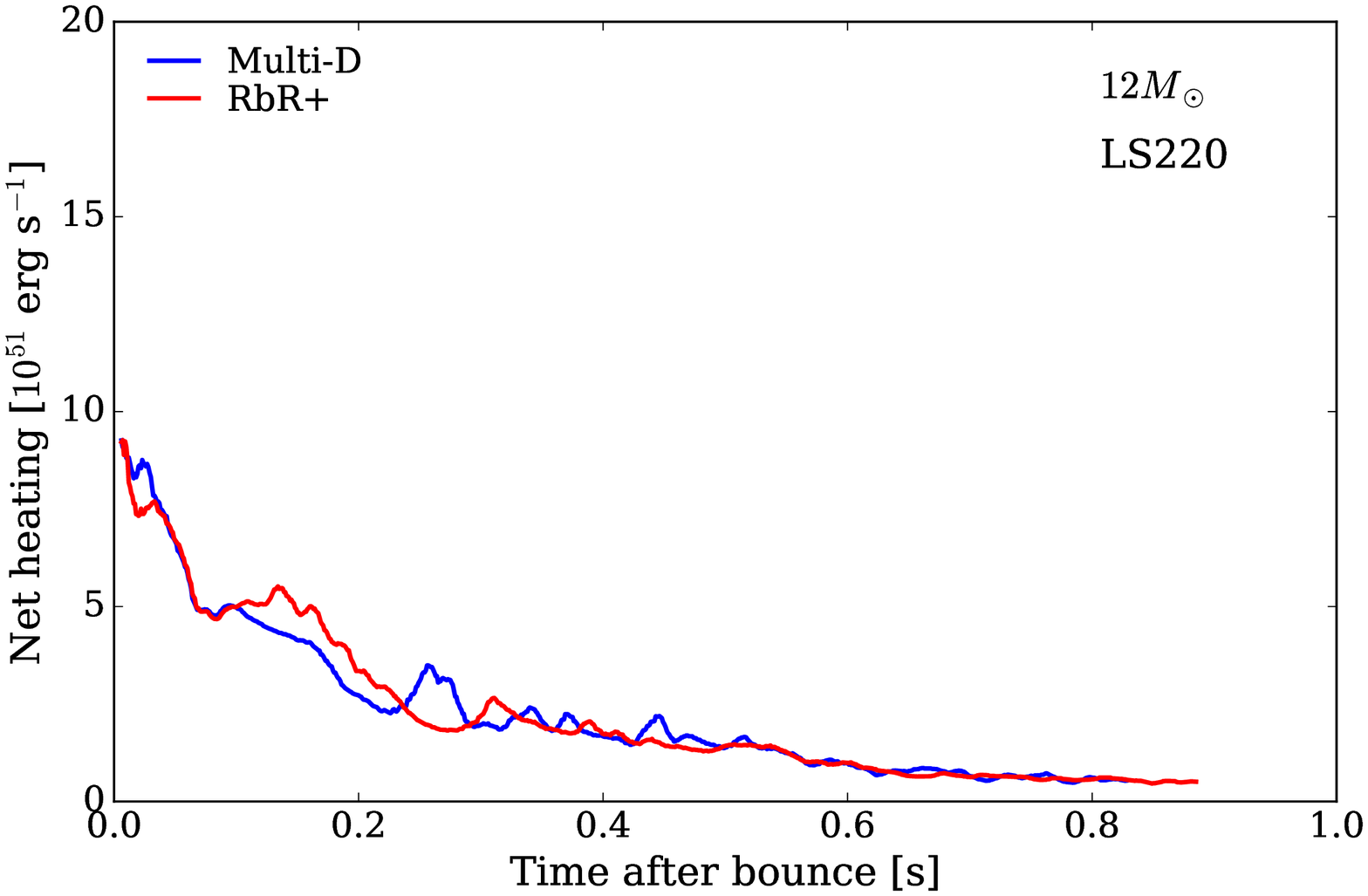}{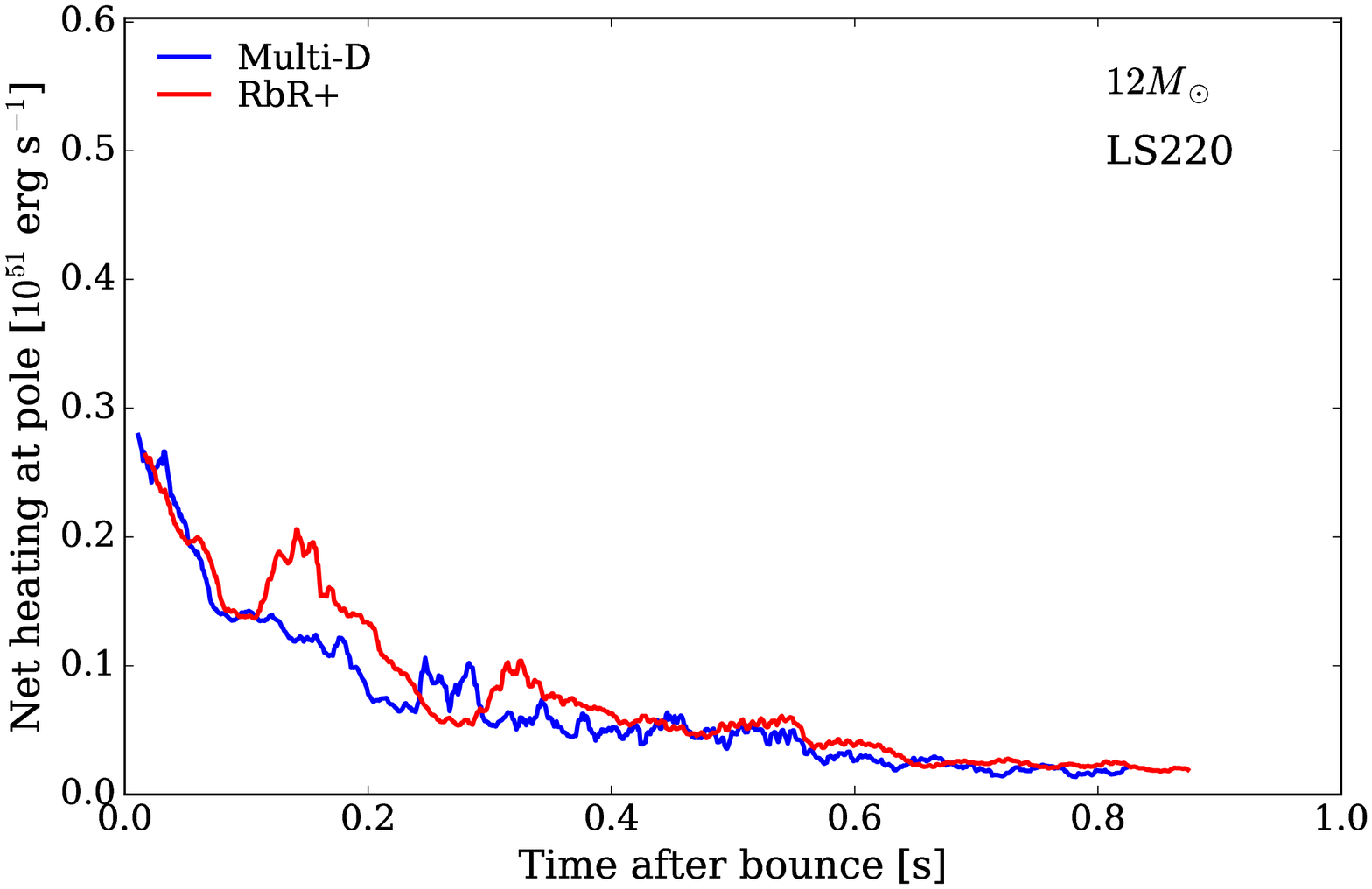}
\caption{{\bf Left:} The integrated neutrino energy deposition, when positive, exterior 
to a radius of 50 kilometers and interior to the shock, for both the multi-D (red) and ray-by-ray+ (blue)
runs of the 12-M$_{\odot}$ progenitor model.  This is roughly the power deposition in the gain region.
{\bf Right:} The same as on the left panel, but exclusively for the polar region 
for $\theta < 20^{\circ}$ (at a correspondingly reduced integrated rate, and not 
``sphericized" as in Fig. \ref{pole1}). Note the difference in scale between the left and the right panels.
For both panels, a 10-millisecond boxcar smoothing convolution was applied.
See the text for discussion.
}
\label{deposition12}
\end{figure}

\begin{figure}
  \centering
  \plottwo{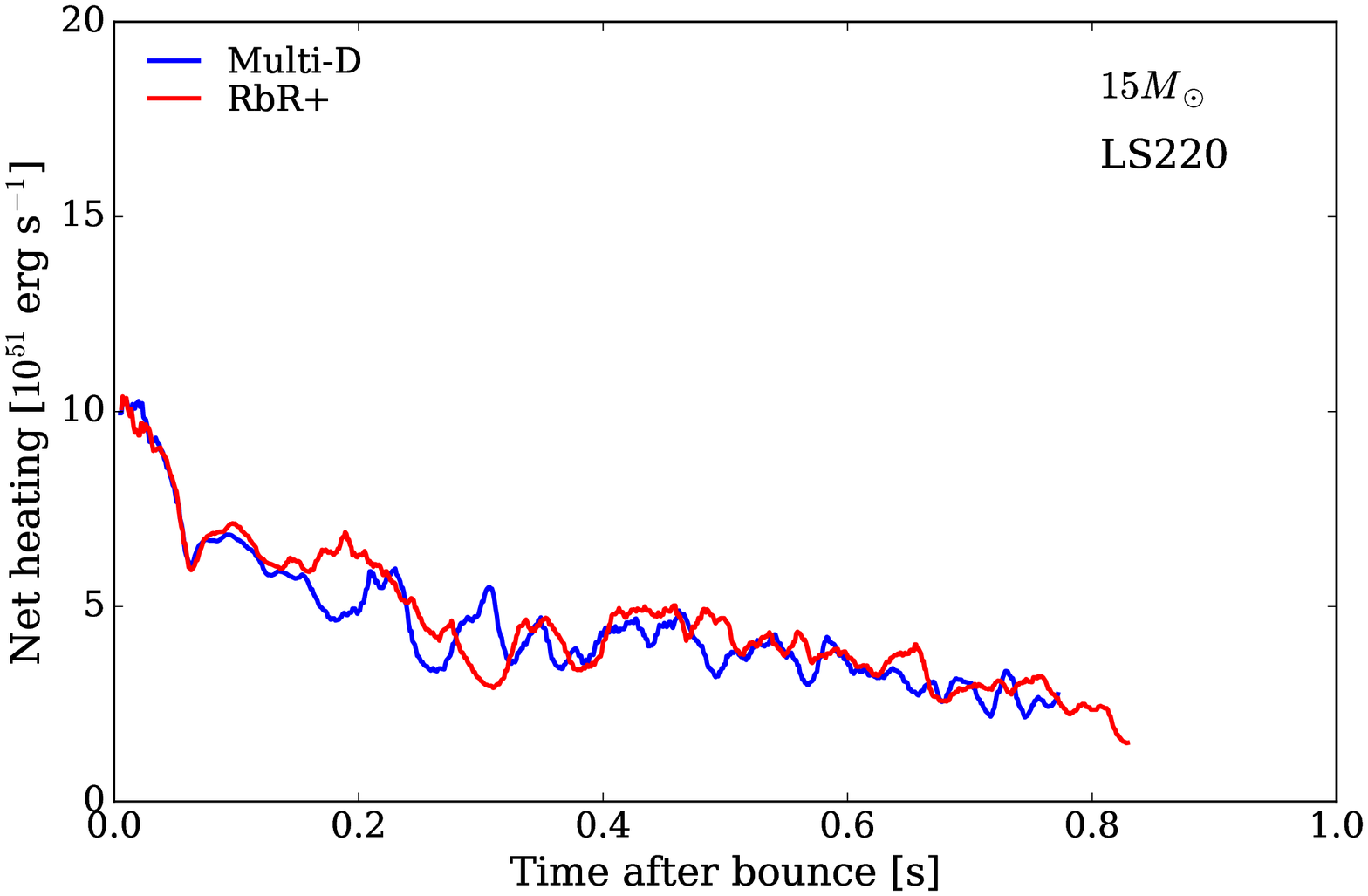}{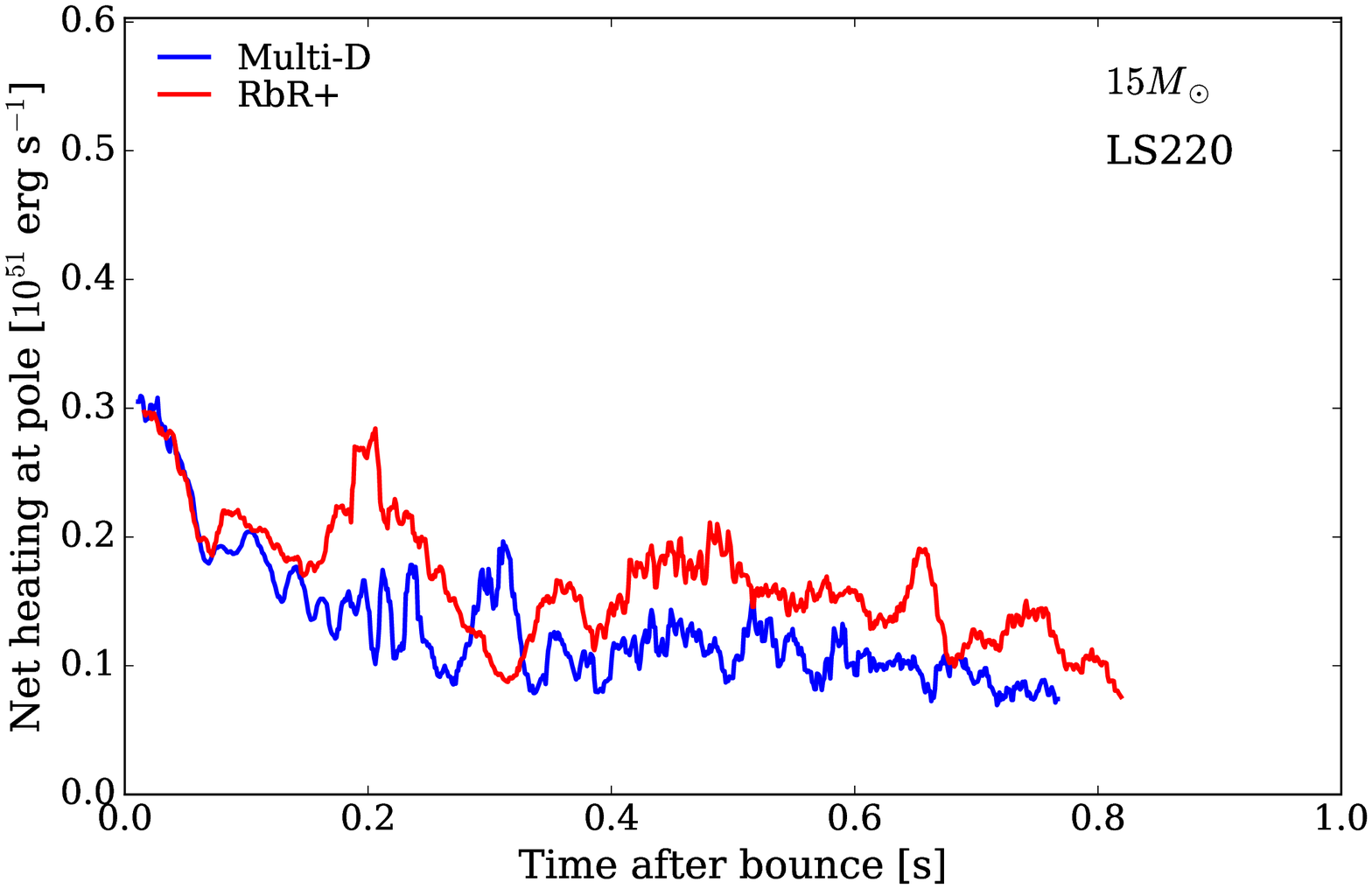}
\caption{The same as Figure \ref{deposition12}, but for the 15-M$_{\odot}$ progenitor model.
}
\label{deposition15}
\end{figure}

\begin{figure}
  \centering
  \plottwo{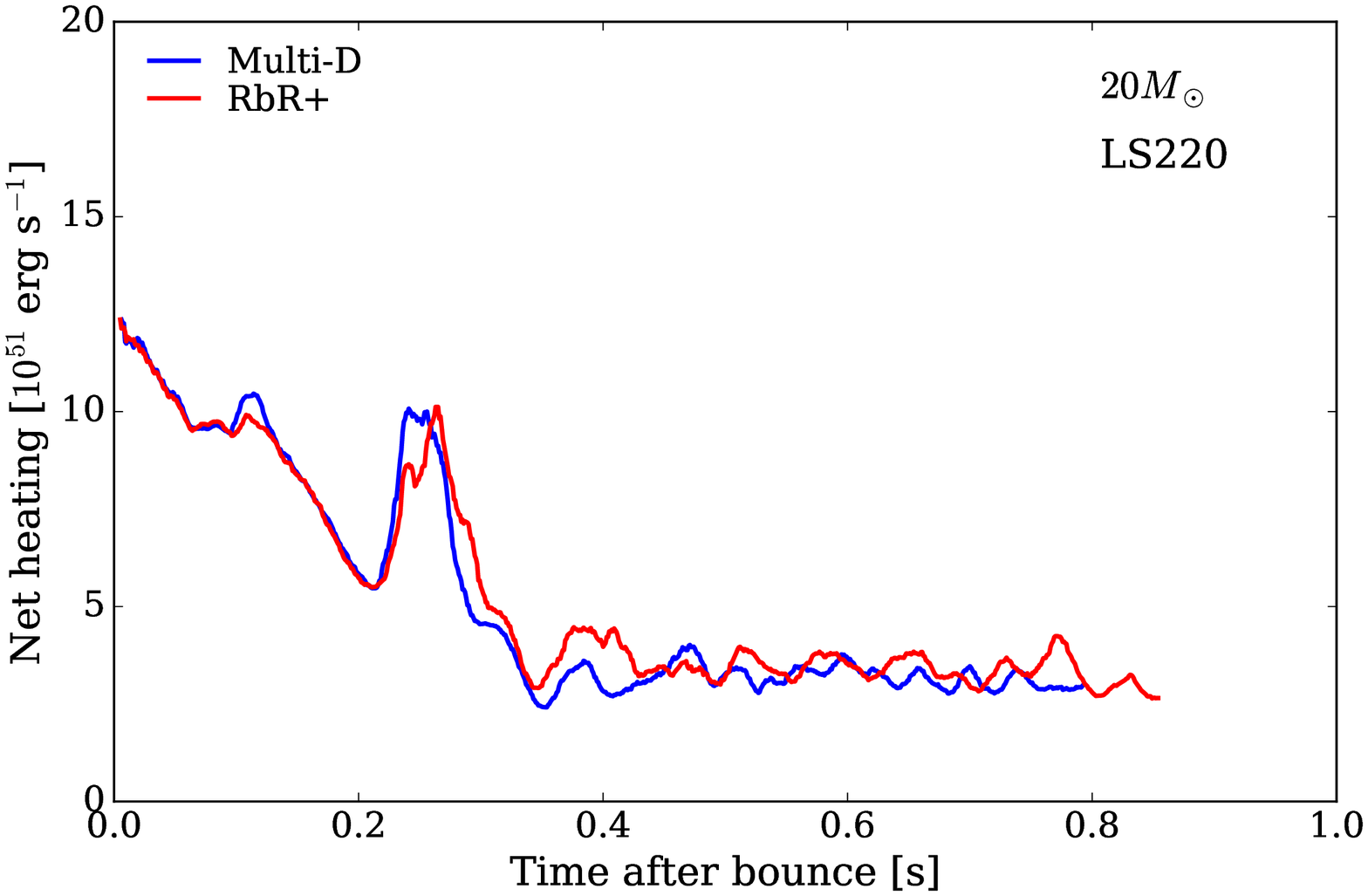}{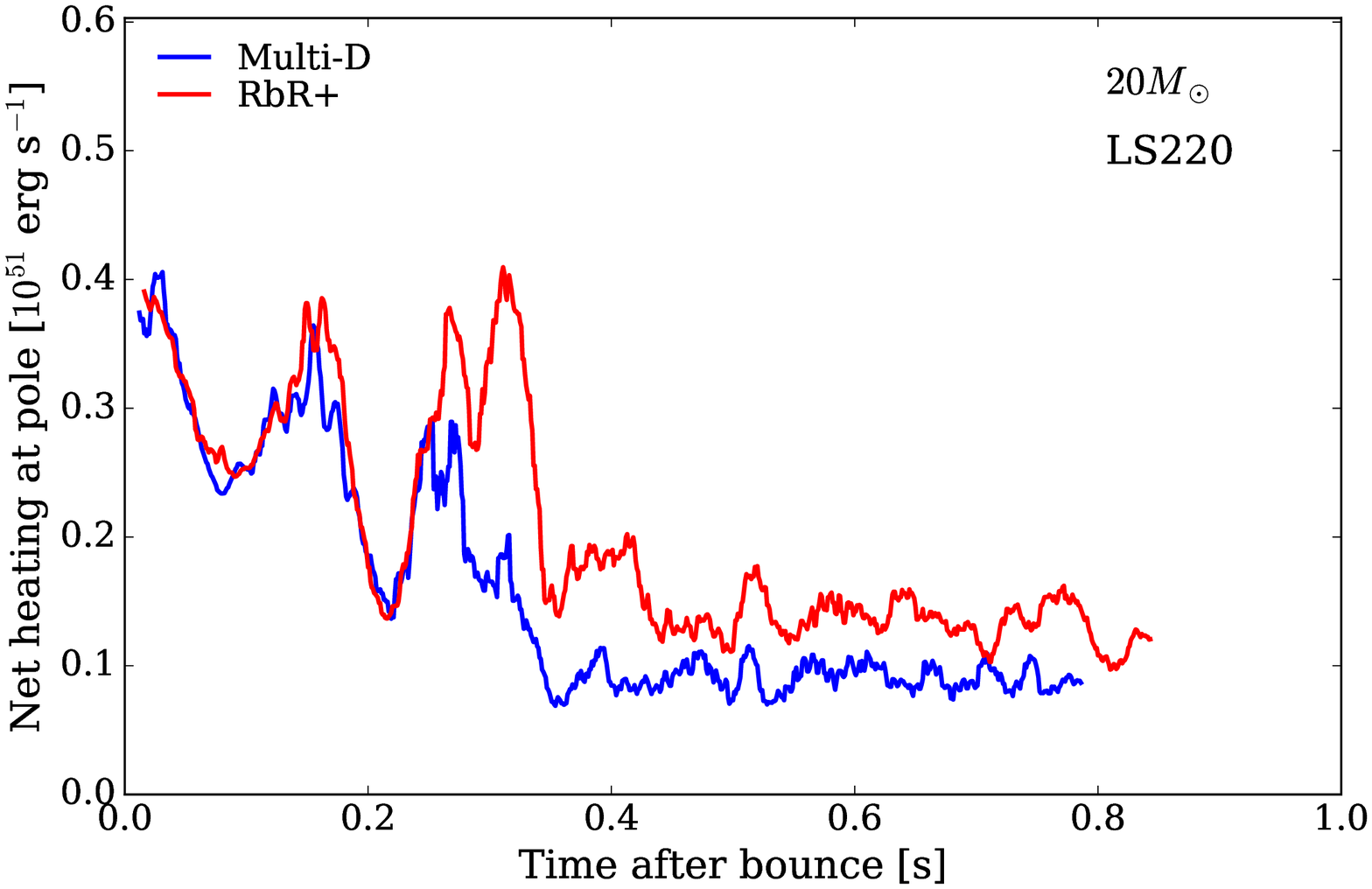}
\caption{The same as Figure \ref{deposition12}, but for the 20-M$_{\odot}$ progenitor model.
}
\label{deposition20}
\end{figure}

\begin{figure}
  \centering
  \plottwo{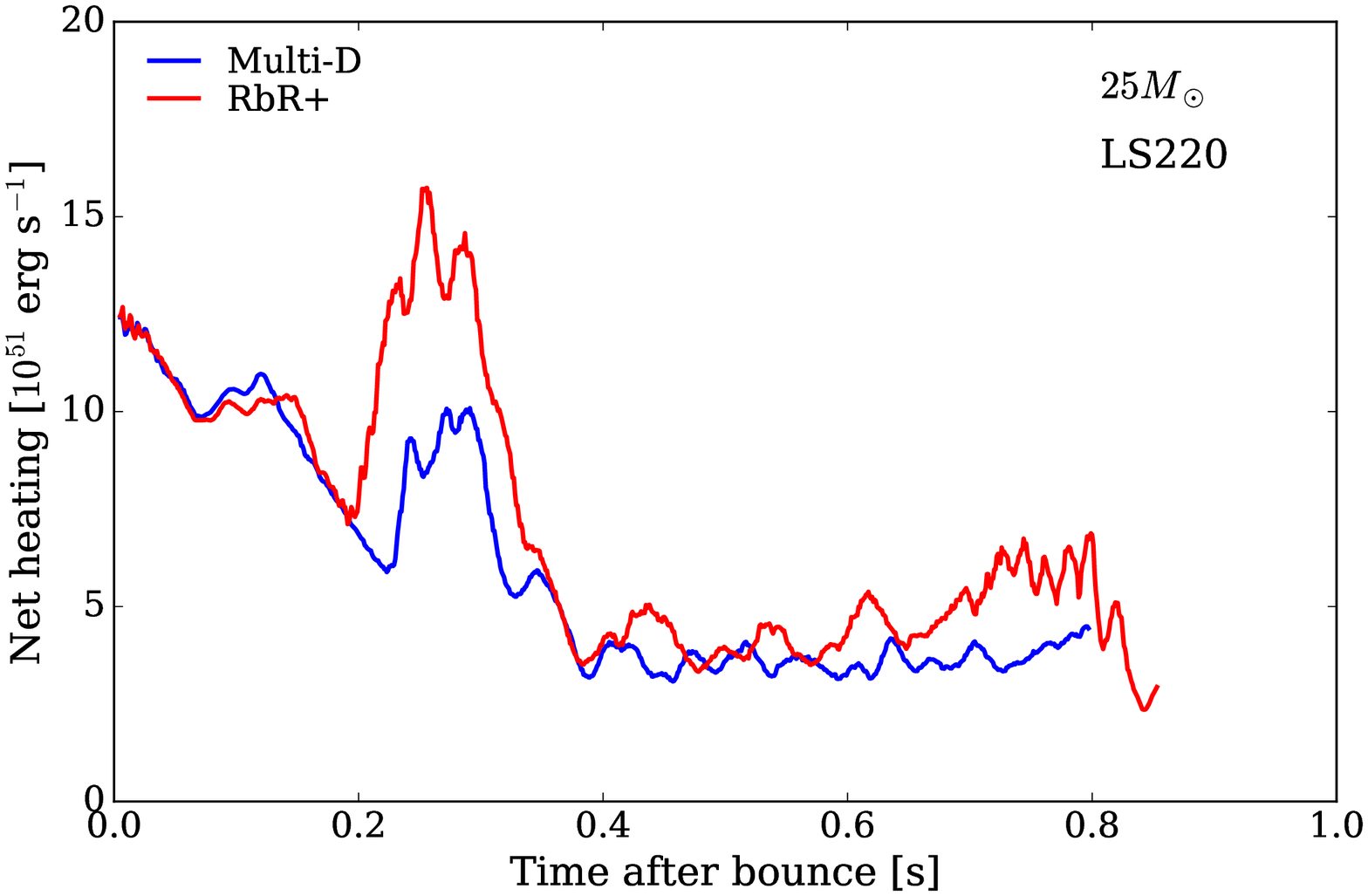}{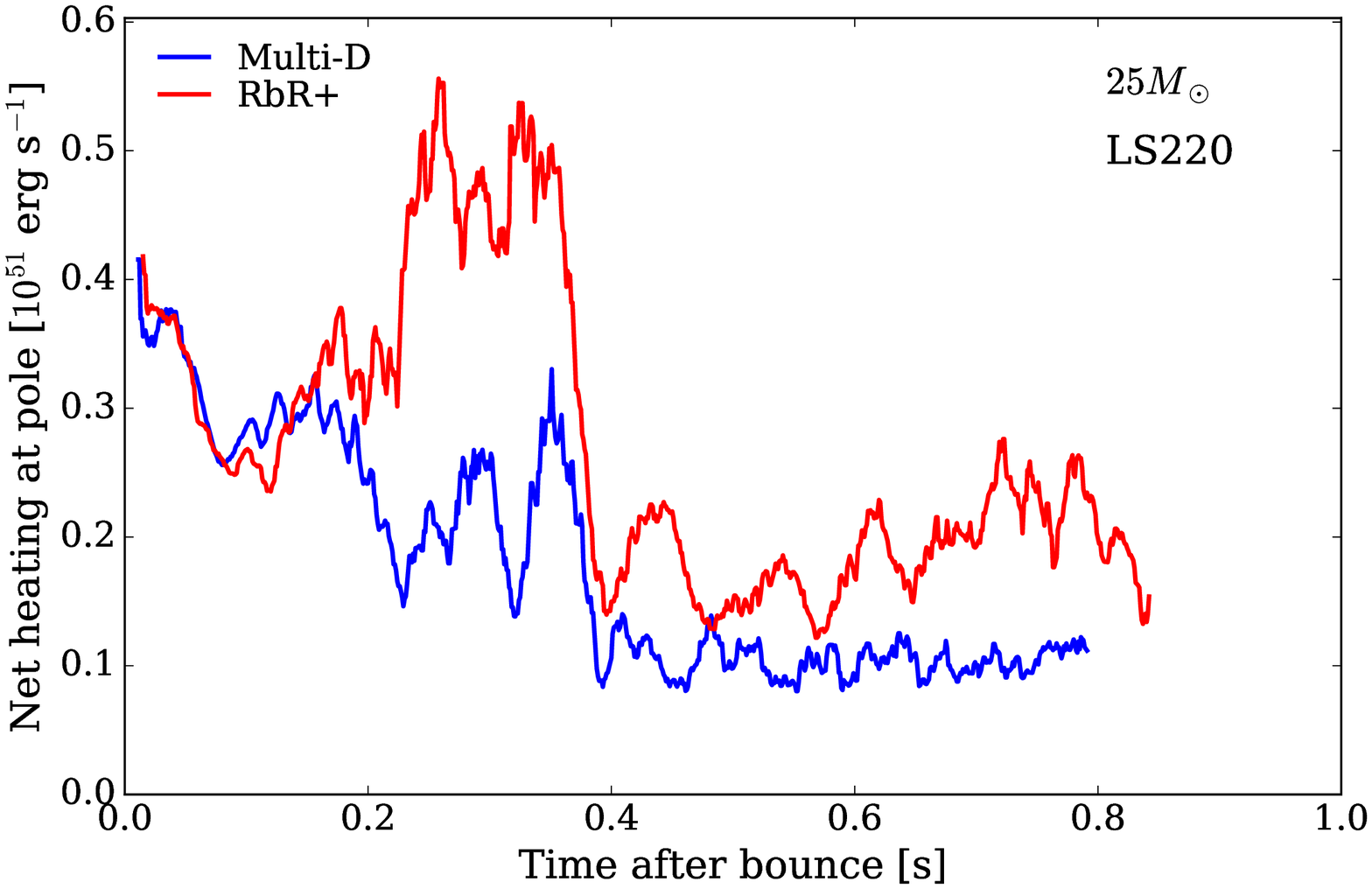}
\caption{The same as Figure \ref{deposition12}, but for the 25-M$_{\odot}$ progenitor model.
}
\label{deposition25}
\end{figure}

Equally important, Figures \ref{deposition12} through \ref{deposition25} provide an estimate of 
the relative degree of neutrino heating in the gain region for this suite of 
four progenitors as a function of time after bounce.  Here, we define the 
gain-region power deposition as the sum of the net $\nu_e$ and $\bar{\nu}_e$ 
heating rates between the shock and the 50-kilometer radius when the net rate 
is positive.  Other definitions can be envisioned, but this measure
is straightforward.  On each panel this quantity is displayed for both the
multi-D and ray-by-ray+ runs.  The left panels provide the total power deposition, 
while the right panels are the same quantity, but for the 20$^{\circ}$ cone around 
the pole. {We see that, for the 12-M$_{\odot}$
model, the differences between the multi-D and ray-by-ray+ are not large, though
in the early post-bounce epoch the ray-by-ray+ numbers can be as 
much as $\sim$50\% higher for the polar quantity.
However, for the other models the differences along the poles 
between those two approaches are progressively more significant with 
increasing progenitor mass, while the total heating rates are not 
much different.  For the 15-M$_{\odot}$, 20-M$_{\odot}$, and 25-M$_{\odot}$
progenitors, the polar power deposition (red) using the ray-by-ray+ prescription can be 
as much as $\sim$10$-$100\% higher than when using the more physical multi-D transport 
approach.}  This behavior is manifest in and parallels the hydrodynamic response, as reflected in
Figure \ref{ray_by_ray.20} in the relative behavior and evolution of the maximum shock radius.    

What Figures \ref{ray_by_ray.12} through \ref{deposition25} collectively seem to 
demonstrate is that the evolution of a 2D model and its neutrino radiation fields can be
quantitatively in error when using the ray-by-ray+ simplification.  Given that the neutrino radiation fields 
and the heating along the poles can be so different, we conclude that the ray-by-ray+ simplification may 
be introducing into the 2D core-collapse supernovae models of those who use 
it an unacceptablely large error. We suggest that the magnitude of these discrepancies is 
at least comparable to those introduced by likely or known uncertainties in the microphysics. 
{Moreover, for our 25-M$_{\odot}$ model the differences between the two transport approaches
translate into a qualitatively different outcome (explosion versus no explosion).}

\subsection{Resolution Dependence}
\label{resol}

The resolution of a simulation can affect a conclusion quantitatively, and the resolution
necessary to demonstrate a given behavior robustly will depend
upon the algorithm employed.  In addition, it has been suggested that 
the ``explodability" of a model seems greater for lower-resolution
simulations (Couch \& Ott 2015).  This is due predominantly to the anomalously greater power
left in such runs at the larger scales at which the turbulent stress is most
significant (Burrows, Hayes, \& Fryxell 1995; Murphy, Dolence, \& Burrows 2013).
Moreover, we surmise that when the initial seed perturbations are not imposed,
the grid resolution will affect both the magnitude and scale of the seed perturbations
for the Rayleigh-Taylor-like instabilities that grow into the turbulence in 
the post-bounce mantles. Smaller-scale perturbations grow more slowly, and in grid codes,
higher-resolution simulations may introduce smaller amplitudes.
When the initial perturbations from which the instabilities grow have 
not been identified and managed (as they have not been in most studies),
then the strength of the resultant turbulence will be similarly contingent.
M\"uller \& Janka (2015) and Couch \& Ott (2015) have started to address 
the importance of the initial turbulent perturbations that arise in the 
pre-collapse convective structures of the cores of massive stars, and we plan 
a separate paper on this phenomenon.  However, here we compare simulations
for ray-by-ray+ and multi-D transport at two different resolutions
to ascertain in a preliminary way the differences.

The 2D explosion simulations of M\"uller, Janka, \& Marek (2012) employed
a spherical polar grid covering 180$^{\circ}$ in $\theta$ with 64 or 128 
angular zones and 10,000 km in radius with 400 radial zones.
They used 12 energy groups from 0 to 380 MeV.
For their 2D explosion simulations, Bruenn et al. (2014) employed
256 uniform angular zones and 512 radial zones out to 23,000 km (25-M$_{\odot}$) 
or 30,000 km (12-M$_{\odot}$), depending upon the progenitor.  
They used 20 energy groups ranging from 4 to 250 MeV.  In neither of these simulations
were the initial perturbations controlled for, and both simulations
employed the ray-by-ray+ approach.  As communicated in \S\ref{code}, the results
described in \S\ref{results} employed 256 angular zones, 608 radial zones 
(covering a radial extent of 10,000 km), and 40 energy groups, and so were rather 
better resolved than the two other efforts.  

\begin{figure}
  \centering
   \plotone{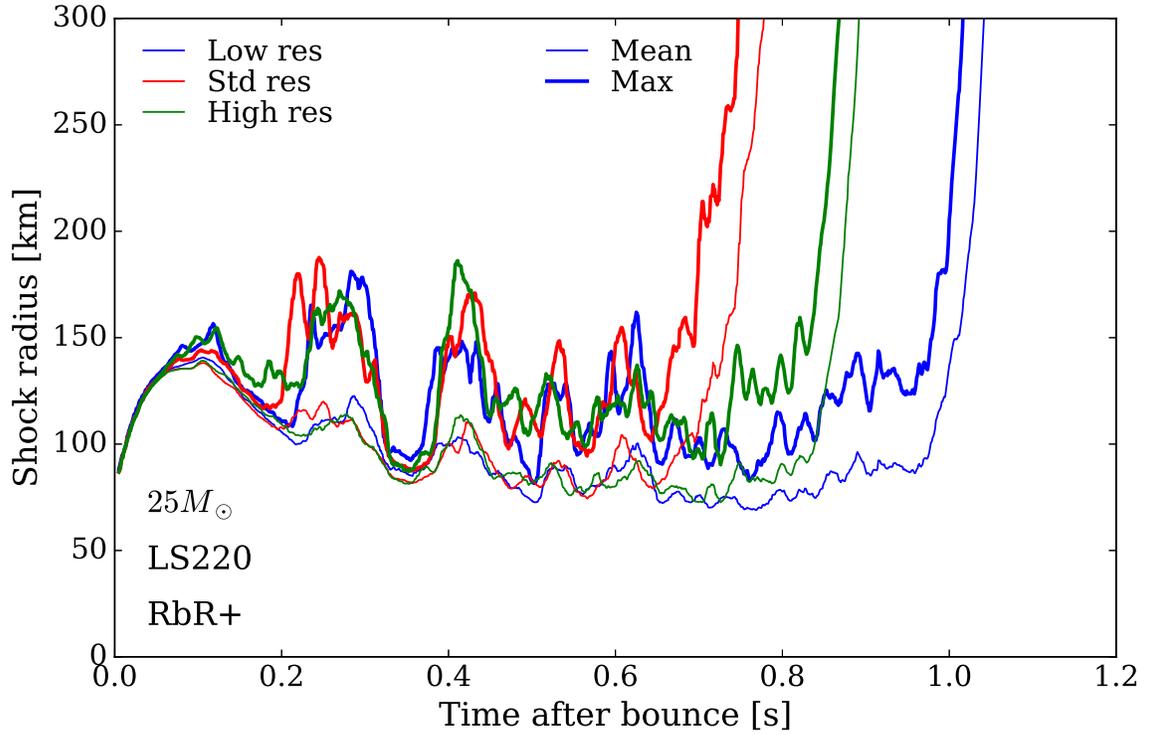}
\caption{This figure provides a comparison for different resolutions of 
the evolution of the shock radius (maximum [thick line] and average/mean [thin line]) of the 25-M$_{\odot}$ progenitor 
ray-by-ray+ model. Low resolution is 304 ($r$) $\times$ 128 ($\theta$) (blue), the baseline resolution 
is 608 $\times$ 256 (red), and the high resolution is 1216 $\times$ 512 (green).
The baseline curves reproduce the data in Figure \ref{ray_by_ray.20} for comparison.
However, in this plot time after bounce goes to 1.2 seconds and the ordinate extends to 300 kilometers.
See the text in \S\ref{resol} for a discussion.
}
\label{newfigure}
\end{figure}

{To gauge the potential effect of resolution, we conducted a set of simulations 
with 512 angular and 1216 radial zones (twice the baseline resolution), 
as well as with 128 angular and 304 radial zones (half the baseline resolution), attempting 
to maintain the aspect ratio of the grid.  Figure \ref{newfigure} shows the comparison 
of the evolution of the shock positions for the ray-by-ray+ runs for the 25-M$_{\odot}$ progenitor at the three 
different resolutions. Despite the chaoticity of the flow and the consequent inherent 
sensitivity to even small changes, all the ray-by-ray+ runs exploded at late times
and manifested similar (though not exactly the same) evolutionary behaviors in R$_{max}$ and R$_{avg}$.
In particular, though many of the oscillations in R$_{max}$ before explosion
are approximately in phase and of similar magnitude, the explosion times for the different
resolutions are not the same, nor are they monotonic with resolution.
Nevertheless, from this we can conclude that the resolution dependence of our runs, at least for the 
25-M$_{\odot}$ progenitor, is minimal and that our simulations are qualitatively
converged.}


\section{Conclusions}
\label{conclusions}

We have presented the first self-consistent, time-dependent calculations 
to address the hydrodynamic consequences of the use in core-collapse simulations 
of the ray-by-ray+ transport ansatz, and our findings 
call into question the wisdom of its pervasive use.
Such a method was introduced to simplify and speed up
otherwise difficult and resource-intensive core-collapse  
simulations, but in so doing it may have compromised the 
results obtained. The combination of ray-by-ray with 
the axial sloshing hydrodynamics that is a feature of 2D supernova dynamics 
(see Burrows, Dolence, \& Murphy 2012, and references therein) can 
result in quantitatively (and, perhaps, qualitatively) incorrect results.  However, we have not
here shown that the same problems arise, or arise to so dramatic a degree,
in 3D. In 3D, the hydrodynamics is very much less prone to axial
motions of the type that signify and define 2D simulations (see, 
for example, Dolence et al. 2013, and references therein).
Also, we have conducted the study using only one code (F{\sc{ornax}}),
incorporating as it does its own algorithms, stencils, and term 
reconstructions. Moreover, there may be further angular, radial, 
or energy-group resolution issues that are manifest to varying degrees in different codes.

However, our study merely compares the same simulations with and without
the ray-by-ray+ simplification. We used the same microphysics, initial models, and 
hydrodynamics, as well as the same resolution (and initial perturbations),
yet we find significant differences of a quantitative nature when doing so.  This conclusion calls into
question simulations in 2D employing the ray-by-ray+ (or simple ray-by-ray)
prescription used by most groups now performing core-collapse simulations
and most groups to date obtaining explosions in 2D by the neutrino 
mechanism.  When there was strong turbulence and large shock radius excursions,
we found that the ray-by-ray+ transport prescription tends to amplify them further. 
{For one progenitor, the 25-M$_{\odot}$, it resulted in explosion, while the full multi-D 
run did not.  We note as far as ``explodability" is concerned that decreasing the energy losses due
to $\nu_{\mu}$ radiation may be as important as increasing the $\nu_{e}$ and $\bar{\nu}_{e}$
heating rates in the gain region. Such decreases could also result from legitimate
alterations in the microphysics and may be one source of the differences in outcome 
seen in the literature.}

{We emphasize that a test of ray-by-ray+ does not require the code 
or calculations to have all the ingredients necessary for a full supernova 
mechanism study; this clearly was not the intention of this paper.  
A clean test merely requires one to compare calculations performed with exactly
the same basic setup and physics, but with either the ray-by-ray+ method or an
actual multi-D transport method employed.  For the ray-by-ray+ method to be acceptable,
such comparison calculations, performed side by side with exactly the same initial
conditions, microphysics and physics, must show little quantitative difference.
However, what we have found is that in such a direct comparison there are
palpable quantitative differences, and occasional qualitative ones.} 

However, from this we do not conclude that the neutrino mechanism
itself is suspect.  It is still a compelling mechanism by which to power
such supernovae.  Rather, aspects of the initial models and the microphysics
might need closer scrutiny, and the behavior in 3D still needs to be 
properly addressed.  The suggestion that sensitivity to given changes 
in the microphysics (for instance, the neutrino-matter coupling) 
increases with increasing dimension (from 1D to 3D) would seem to be 
a productive line to follow (Melson et al. 2015).  Differences between
the initial 1D progenitor models currently employed
in core-collapse work and those that better incorporate 1) convection 
and semi-convection (probably fully 3D initial models; Couch et al. 
2015; Arnett et al. 2015), 2) initial convective perturbations (M\"uller 
\& Janka 2015; Couch \& Ott 2015), and 3) rotation (Ott et al. 2006,2008) are certainly 
suggestive and demand a closer look.  As we have here suggested, the initial perturbation
structure has not been adequately scrutinized.

Whatever the ultimate solution to the long-standing problem of the 
mechanism of core-collapse supernova explosions, our study 
serves once again to emphasize that it is crucial 
to ensure the physical fidelity of numerical methods that are applied 
to multi-dimensional problems in which turbulence and chaotic flow are 
of qualitative import. The next step is for other groups to check what we
have found in 2D and to test this result in the 3D context.  In fact, 
there is a rather long list of numerical challenges and code verification 
issues yet to be met collectively by the world's supernova modelers.  The results
of different groups are still too far apart to lend ultimate credibility
to any one of them.  Perhaps our study can serve to encourage the necessary 
global code verification initiative.


\acknowledgments

The authors acknowledge the help of Evan O'Connor with the Lattimer-Swesty 
equation of state and numerous useful conversations with Christian Ott. 
{We also thank the anonymous referee for encouraging us to further 
minimize the numerical noise in the dendritic region of the grid.} 
Support was provided by 
the NSF PetaApps program, under award OCI-0905046 via a subaward
no. 44592 from Louisiana State University to Princeton University and
by the Max-Planck/Princeton Center (MPPC) for Plasma Physics (NSF PHY-1144374).
The authors employed computational resources provided by the TIGRESS
high performance computer center at Princeton University, which is jointly supported by the Princeton
Institute for Computational Science and Engineering (PICSciE) and the Princeton University Office of
Information Technology, and by the National Energy Research Scientific Computing Center
(NERSC), which is supported by the Office of Science of the US Department of
Energy under contract DE-AC03-76SF00098. 
This work is part of the ``Three Dimensional Modeling of Core-Collapse      
Supernovae" PRAC allocation support by the National Science Foundation (award number ACI-1440032) 
and is part of the Blue Waters sustained-petascale computing project, 
which is supported by the National Science Foundation 
(awards OCI-0725070 and ACI-1238993) and the state of Illinois. 
Blue Waters is a joint effort of the University of Illinois at 
Urbana-Champaign and its National Center for Supercomputing Applications.
%
This paper has been assigned a LANL preprint \# LA-UR-15-28756.

\begin{appendix}

The basic comoving-frame equations of radiative transfer that we solve are the zeroth and first moment 
equations of the full equation of radiative transfer for the specific intensity of 
the $\nu_e$, $\bar{\nu}_e$, and $\nu_{\mu}$ neutrinos, where the latter represents 
the $\nu_{\mu}$, $\bar{\nu}_{\mu}$, $\nu_{\tau}$, and $\bar{\nu}_{\tau}$ neutrinos collectively. The equations are:

\begin{align}
E_{s\varepsilon,t} + (F_{s\varepsilon}^i + {\bf v^i E_{s\varepsilon}})_{;i} - {\bf v^i_{;j}\frac{\partial}{\partial\ln\varepsilon} P_{s\varepsilon i}^j} &= j_{s\varepsilon} - c \kappa_{s\varepsilon} E_{s\varepsilon} \\
F_{s\varepsilon j,t} + (c^2 P_{s\varepsilon j}^i + {\bf v^i F_{s\varepsilon j}})_{;i} + {\bf v^i_{;j} F_{s\varepsilon i} - v^i_{;k} \frac{\partial}{\partial\varepsilon} (\varepsilon Q^k_{s\varepsilon ji})} &= -c(\kappa_{s\varepsilon} + \sigma^{\rm tr}_{s\varepsilon}) F_{s\varepsilon j}\, ,
\end{align}
where differentiation is indicated with standard notation, $\varepsilon$ is the neutrino energy, $s\in\{\nu_e,\bar{\nu}_e,\nu_{\mu}\}$,
$E_{s\varepsilon}$ is the radiation energy density spectrum (zeroth moment), $F_{s\varepsilon j}$ is radiation flux 
spectrum (first moment),  $P_{s\varepsilon i}^j$ is the radiation pressure tensor (second moment), $Q^k_{s\varepsilon ji}$ 
is the heat tensor (third moment), $\sigma^{\rm tr}_{s\varepsilon}$ is the transport 
scattering opacity, and the other variables have their standard meanings. 
We use the M1 closure to truncate the radiation moment hierarchy by specifying the second and third 
moments in terms of the zeroth and first moments.  

For the ray-by-ray+ solutions, the transverse components of the radiation 
flux are set to zero, while we retain the velocity advection
terms, even the transverse advection terms (all shown in bold above). {Of course, for
both schemes we retain the advection of Y$_e$ in all directions.  In our 
implementation of ray-by-ray+, the momentum exchange with the matter due to any 
lateral neutrino stress is not incorporated (though the corresponding radial stress is), since 
these lateral force terms are very small. This is particularly true in the semi-transparent region 
where the overall differences between ray-by-ray+ and 2D transport are largest and most relevant.  
In the Buras et al. (2006) implementation of ray-by-ray+, the authors include a transverse 
diffusive term that approximates the lateral neutrino force above densities 
of 10$^{12}$ g cm$^{-3}$ by a gradient term in the neutrino energy density 
(Eddington approximation), setting it to zero below this density.  We feel that 
including such a term in our ray-by-ray+ implementation is not only unnecessary, but 
too approximate and ad hoc. Moreover, including such an effect is not relevant 
when researching the differences between the two methodologies in the  
shock dynamics and in the hydrodynamic outcomes.}

\end{appendix}


\newpage

\end{document}